\documentclass[%
reprint,aps,prc,
superscriptaddress,
showpacs, preprintnumbers,
nofootinbib,    
amsmath,
amssymb,
]{revtex4-2}

\usepackage{graphicx} 
\usepackage{dcolumn}
\usepackage{bm}
\usepackage{lmodern} 
\usepackage{amssymb}
\usepackage{gensymb}
\usepackage{caption}
\usepackage{sidecap}
\usepackage{color}
\usepackage{float}
\usepackage{multirow}

\usepackage[font=footnotesize]{caption}
\usepackage{lineno}
\usepackage{soul}

\newcommand\scalemath[2]{\scalebox{#1}{\mbox{\ensuremath{\displaystyle #2}}}}
\newcommand{\liseppmath}{LISE{$\scalemath{0.7}{^{^{++}}}$}}

\newcommand{\softsh}[1]{\texttt{#1}}
\newcommand{\soft}[1]{\texttt{#1}$\,$}

\newcommand{\lisepp}{\soft{\liseppmath}}

\newcommand{\nucl}[2]{\hbox{$^{#1}$#2}}
\newcommand{\Aq}{$A$/$q$ }
\newcommand{\Aqsh}{$A$/$q$}
\newcommand{\EER}{\soft{3EER }}
\newcommand{\EERsh}{\softsh{3EER}}

\begin{document}

\preprint{APS. Version 2.3}

\title{Abrasion-fission reactions at intermediate energies}

\newcommand{\aNSCL}{\affiliation{National Superconducting Cyclotron Laboratory, Michigan State University, East Lansing, MI 48824, USA }}
\newcommand{\aMSUphys}{\affiliation{Department of Physics and Astronomy, Michigan State University, East Lansing, MI 48824, USA}}
\newcommand{\aFRIB}{\affiliation{Facility for Rare Isotope Beams, Michigan State University, East Lansing, MI 48824, USA}}

\author{M.~Bowry}
\altaffiliation{Present address: School of Computing, Engineering and Physical
Sciences, University of the West of Scotland, High Street, Paisley PA1 2BE, United Kingdom\\  Michael.Bowry@uws.ac.uk}
\aNSCL

\author{O.~B.~Tarasov}  \aFRIB

\author{J.~S.~Berryman} \aFRIB
\author{V.~Bader} \aNSCL

\author{D.~Bazin} \aFRIB \aMSUphys

\author{T.~Chupp} \affiliation{Department of Physics, University of Michigan, Ann Arbor, Michigan 48104, USA}
\author{H.L.~Crawford} \affiliation{Nuclear Science Division, Lawrence Berkeley National Laboratory, Berkeley, California 94720, USA}

\author{A.~Gade} \aFRIB \aMSUphys

\author{E.~Lunderberg} \aNSCL

\author{A.~Ratkiewicz}
\altaffiliation{Present address: Lawrence Livermore National Laboratory, Livermore, California 94550, USA}
\affiliation{Department of Physics and Astronomy, Rutgers University, New Brunswick, New Jersey 08903, USA}

\author{F.~Recchia}
\altaffiliation{Present address: Dipartimento di Fisica e Astronomia dell’Universita and INFN, Sezione di Padova, I-35131 Padova, Italy}
\aNSCL

\author{B.~M.~Sherrill} \aFRIB \aMSUphys
\author{D.~Smalley}  \aNSCL
\author{A.~Stolz} \aFRIB \aMSUphys

\author{S.~R.~Stroberg}
\altaffiliation{Present address: Department of Physics, University of Notre Dame, Notre Dame, Indiana 46556, USA}
\aNSCL

\author{D.~Weisshaar} \aFRIB
\author{S.~Williams} \aNSCL

\author{K.~Wimmer}
\altaffiliation{Present address: GSI Helmholtzzentrum f\''ur Schwerionenforschung, Darmstadt, Germany}
\aNSCL

\author{J.~Yurkon}\aNSCL

\date{\today}

\begin{abstract}

The availability of high-intensity, heavy-ion beams coupled to sensitive, large solid-angle-acceptance spectrometers has enabled a detailed examination of the fission fragments produced in induced-fission reactions. The \textit{abrasion-fission} process involves the formation of projectile-like prefragments in violent nuclear collisions at relative energies in excess of 100~MeV/u. At intermediate energies below this threshold, experiments suggest a change in the prefragment kinematic qualities. Information regarding the influence of this transitional phase upon the evolution of nuclei approaching the point of scission is scarce. In this article, data are presented for over 200 nuclei from nickel to palladium produced in abrasion-fission reactions of a 80~MeV/u $^{238}$U beam. Cross sections were obtained following yield measurements performed for the principal charge states of the identified fission fragments and a detailed analysis of the ion transmission. A full kinematic analysis of the fission fragments has been performed using the \lisepp software package, where the trajectory of an ion passing through a spectrometer can be reconstructed based upon measurements at the focal plane. The results obtained at the S800 spectrograph are compared  with predictions obtained with a three-fission progenitor (\EERsh) model. Systematic studies of fission-fragment properties continue to provide a valuable experimental benchmark for theoretical efforts directed toward describing this complex decay channel, that is important in the context  of planning experiments to explore the neutron-rich region of the nuclear chart at rare-isotope beam facilities.

\begin{description}
\item[PACS numbers]
29.30.Kv, 25.75.-q, 27.80.+w
\end{description}
\end{abstract}

\maketitle

\section{\label{level1}Introduction}
Fission decay is among the most well-known manifestations of the nuclear many-body problem where the interplay between the liquid-drop approximation and the nuclear shell model both offer insight into the behavior of the fissioning system toward the point of scission. A microscopic description of the process remains very challenging, not the least due to the dynamic and complex evolution of the single-particle states in the many-body system during the deformation of the nuclear fluid. Scientific interest in fission decay products ranges from fundamental nuclear structure measurements and nuclear astrophysics \cite{Goriely13} to direct applications such as constraining sources of decay heat in nuclear reactors \cite{Algora10}.\par

The introduction of intense beams of accelerated heavy ions enabled access to a different type of induced fission reaction at relativistic velocities, so-called abrasion-fission reactions. Nucleons removed from the projectile beam following collisions with target nuclei leave a projectile-like prefragement in a highly excited state \cite{Schmidt93}, decaying in-flight via fission and/or through the evaporation of light particles ($p$,$n$,$\alpha$) and $\gamma$ rays. Electromagnetically-induced (i.e. Coulomb) fission is also possible when high-$Z$ targets are used.

Important work in this area was performed during the 1990s and 2000s using the in-flight Fragment Separator FRS at the Gesellschaft f\"{u}r Schwerionenforschung GSI (Darmstadt, Germany) \cite{ Bernas97, Enqvist99, Schmidt01, Bernas03, Pereira07} where uranium beams with energies of 1~GeV/u were available. The majority of these studies have focused on the production cross sections and velocity distributions of fission fragments as a function of $N$ and $Z$. A dependence of the fission fragment mass distributions upon excitation energy was observed giving rise to both asymmetric and symmetric mass splitting. Asymmetric distributions result from low-energy Coulomb- or abrasion-fission, the latter corresponding to the removal of a small number of nucleons. In this domain, the magic nucleon numbers greatly influence the $N$, $Z$ and kinetic energy of the fission fragments. Other phenomena include (i) an increase in the total Coulomb-fission cross section for higher beam energies \cite{Enqvist99}, (ii) the influence of deformation on the resulting kinetic energy (from Coulomb repulsion) of the most neutron-rich fission fragments \cite{Pereira07}, and (iii) the influence of the deformation energy upon the post-scission neutron flux.\par

The instability of a given nucleus to fission is broadly defined by the fissility parameter, equal to $Z^{2}/A$ \cite{Cohen74}, where the most neutron-deficient species are more likely to fission than their neutron-rich counterparts for a given~$Z$. The range of possible ``parent'' nuclei was previously estimated for the reactions $d$\:($^{238}$U,\:X) and Pb\:($^{238}$U,\:X)~\cite{Pereira07,Enqvist99} by assuming an unchanged charge density between the parent and daughter nuclei and comparing the measured fission fragment velocities with the available kinetic energy calculated for different parent $Z$.\par
At relative energies of several hundreds of MeV per nucleon, the symmetric mass distribution is favored and accounts for the majority of the in-flight fission cross section, typically of the order of a few barns \cite{Enqvist99}. Mass distributions for different isotopic chains were observed to obey Gaussian shapes although an enhancement in the production of the most neutron-rich species for a given $Z$ (particularly for high-$Z$ fragments) was observed and attributed to Coulomb-fission of the heaviest, most neutron-rich projectile prefragments.\par

It is expected that the in-flight fission products may also provide an additional probe of the initial relativistic collisions. At intermediate energies, a reduction in the prefragment momentum width $\sigma_{0}$~\cite{Goldhaber74} is observed for $E/A \approx$~30-100~MeV/u~\cite{OT-NPA04}. Experimental data regarding abrasion-fission at low and intermediate energies are limited to in-flight fission reactions such as Be, Pb($^{238}$U, X) performed at 345~MeV/u at the RIKEN Nishina Center Radioactive Ion Beam Factory (Japan) \citep{Ohnishi08,Ohnishi10,Suzuki13,Fukuda18} and fusion-/transfer-induced fission at the Grand Acc\'{e}l\'{e}rateur National d'lons Lourds (GANIL, France) via $^{12}$C($^{238}$U, X) at 6~MeV/u \cite{Caamano13}, and via Be, C($^{238}$U, X) at 24~MeV/u~\cite{OT-EPJA18}. Fission fragments were also produced in standard kinematics via the fragmentation of uranium targets using light-ion beams \citep{McGaughey85} where cross sections were estimated following $\gamma$-ray spectroscopy of the irradiated targets. 

The data presented in the current work, therefore, provides a unique perspective on the study of fission reaction mechanisms and fisson fragment properties in inverse kinematics at energies relevant to the transitional region between $E$~$\approx$~30-100~MeV/u.
A previous study at MSU with a uranium beam at 80~MeV/u was aimed at a search for new microsecond isomers in neutron-rich nuclei~\cite{Folden09}. Studying the abrasion-fission mechanism at intermediate beam energies is important for planning experiments to produce the most neutron-rich nuclei near the neutron drip line \citep{Neufcourt20} at the new Facility for Rare Isotope Beams (FRIB) \cite{Glasmacher17, Sherrill18}.

\subsection{\label{level2_AF}Theoretical treatment of abrasion-fission reactions}

The calculation of cross sections in abrasion-fission reactions is performed in two stages. In the first stage of the reaction (abrasion), the excited prefragments are formed, and de-excite in the second stage (ablation) via the emission of light particles, intermediate-mass fragments and 
fission. The \soft{ABRABLA} Monte Carlo code was previously used to calculate cross sections and velocities of residues produced in relativistic heavy-ion collisions and has demonstrated good predictive power for reactions with a \nucl{238}{U} beam  \cite{Kurcewicz12}. \soft{ABRABLA} includes an improved version of the abrasion model for peripheral and mid-peripheral collisions of relativistic heavy ions \cite{Gaimard91} and a statistical de-excitation model \cite{ABLA07} to compute each stage of the abrasion-fission reaction, respectively.

\begin{figure*}
\includegraphics[width=\textwidth]{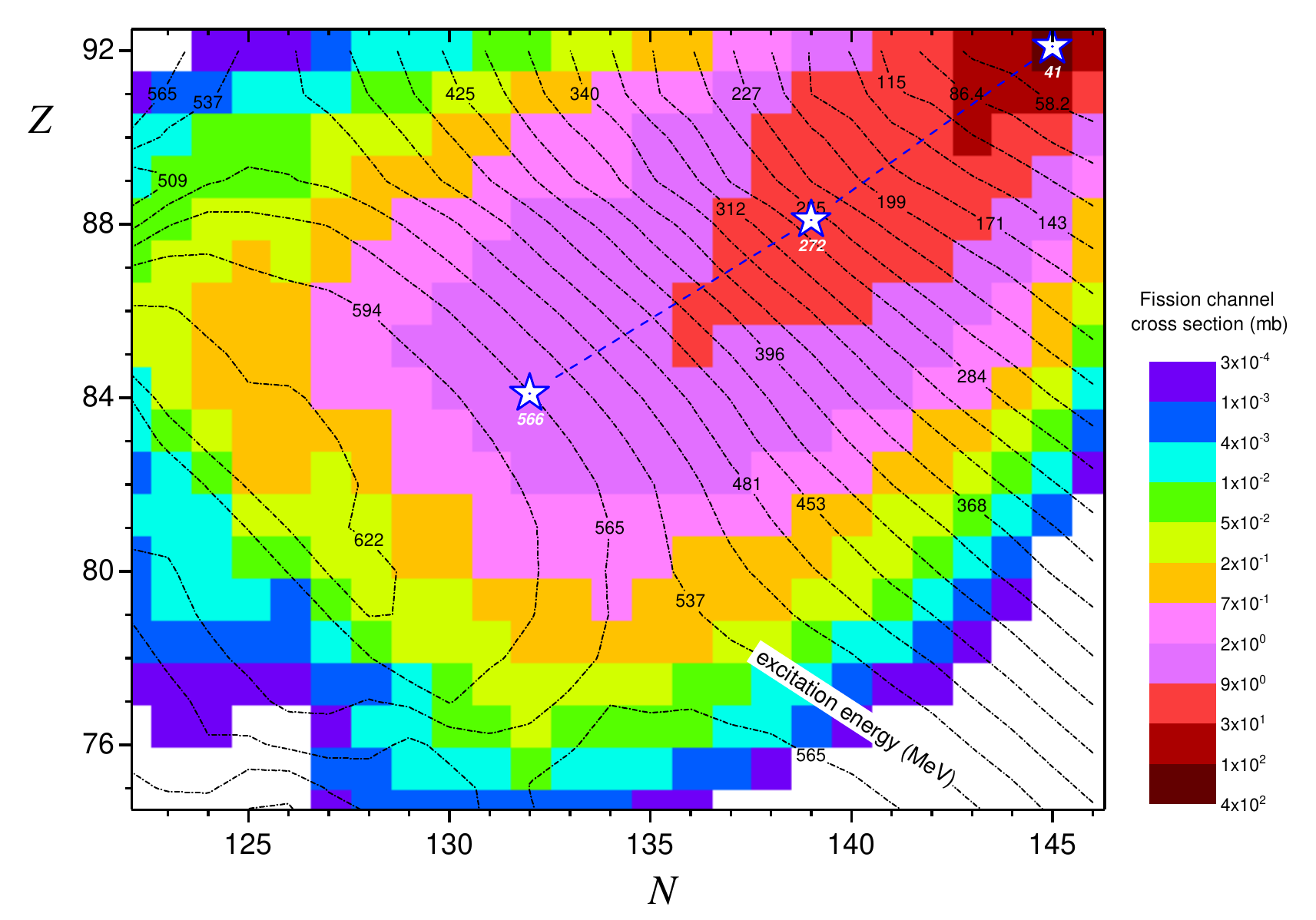}
\caption{Map of nuclei formed following abrasion in the interaction of $^{238}$U + $^{12}$C calculated with the \lisepp three excitation energy regions (\EERsh) model. 
 The contour lines denote the excitation energies of nuclei assuming 27~MeV per abraded nucleon. Stars show the EER nuclei used in this work.}
\label{fig:FissileNucleiMap}
\end{figure*}

While the Monte Carlo method may provide an estimate of the cross sections, calculations typically require a significant amount of time to complete and this may be amplified by the necessity to perform calculations for many nuclei and/or exotic nuclei with low a probability of production. In addition, there is a lack of flexibility in optimizing the model parameters when comparing results with experimental data.
An alternative approach is presented here in the form of an analytical model, \EER~\cite{LISE_AF_preprint,LISE_AF}, developed within the \lisepp framework \cite{Tarasov08}, where some simplifications are used.

For example, the description of the ablation stage, which is based on 
the transport integral theory of Ref.~\cite{Bazin94}, does not take into 
account the angular momentum of the prefragments. While angular momentum 
is very important for the modeling of multi-nucleon transfer reactions 
at low energies, the sensitivity of the residues produced in high-energy 
abrasion-type reactions is expected to be lower as numerical 
calculations have shown that mainly lower angular momenta, around 
10$\,\hbar$ to 20$\,\hbar$ combined with large excitation energies, are 
involved \cite{deJong97}.

The \EER model selects three nuclei to reproduce a complete abrasion-fission yield. The three fission progenitors correspond to prefragments formed with different excitation energies (EER, excitation energy regions) which are proportional to the number of nucleons abraded in the initial collision. These nuclei are selected based on the results of a first-stage abrasion calculation, shown in Figure~\ref{fig:FissileNucleiMap}. In the initial approximation, when describing all progenitor nuclei, the boundaries are selected in such a way that the cross-sections are evenly divided between the regions. Progenitor excitation energies, proton numbers, and neutron numbers are suggested to be cross-section weighted averages. The parameters used to define the boundaries of the excitation energy regions and the production cross sections of the selected nuclei are given in Table~\ref{tab:eEER_param}. During development of the model it was observed that the ability to reproduce the yields of high-Z ($\approx$~70), neutron-rich fragments is closely correlated with the selection of the progenitor nuclei. In particular, an improvement was obtained by selecting highly-excited prefragments with a neutron number close to that of the projectile.

\begin{table}[h]
\begin{center}
\caption{Summary of the excitation energy region (EER) parameters used in the current work.}
\label{tab:eEER_param}
\setlength{\tabcolsep}{3pt} 
\renewcommand{\arraystretch}{1.2} 
\begin{tabular}{|c|rcl|c|c|c|}
\hline
            & \multicolumn{3}{c|}{Region}       &           & Excitation    & Cross   \\
EER         & \multicolumn{3}{c|}{boundaries}   &  Nucleus  &    energy     &  section             \\
            & \multicolumn{3}{c|}{(MeV)}        &           & (MeV)         & (mb)                  \\

\hline
Low        &                      & $E_{x}$  &   $\le 83 $      &$^{237}$U    & 41 &  503\\
Medium   &         $83 <$   & $E_{x}$  &    $\le 330$    & $^{227}$Ra & 272 & 676\\
High        &         $330 <$ & $E_{x} $ &                      & $^{216}$Po  & 566 &  489\\
\hline
\end{tabular}
\end{center}
\end{table}

\begin{figure}
\includegraphics[width=\columnwidth]{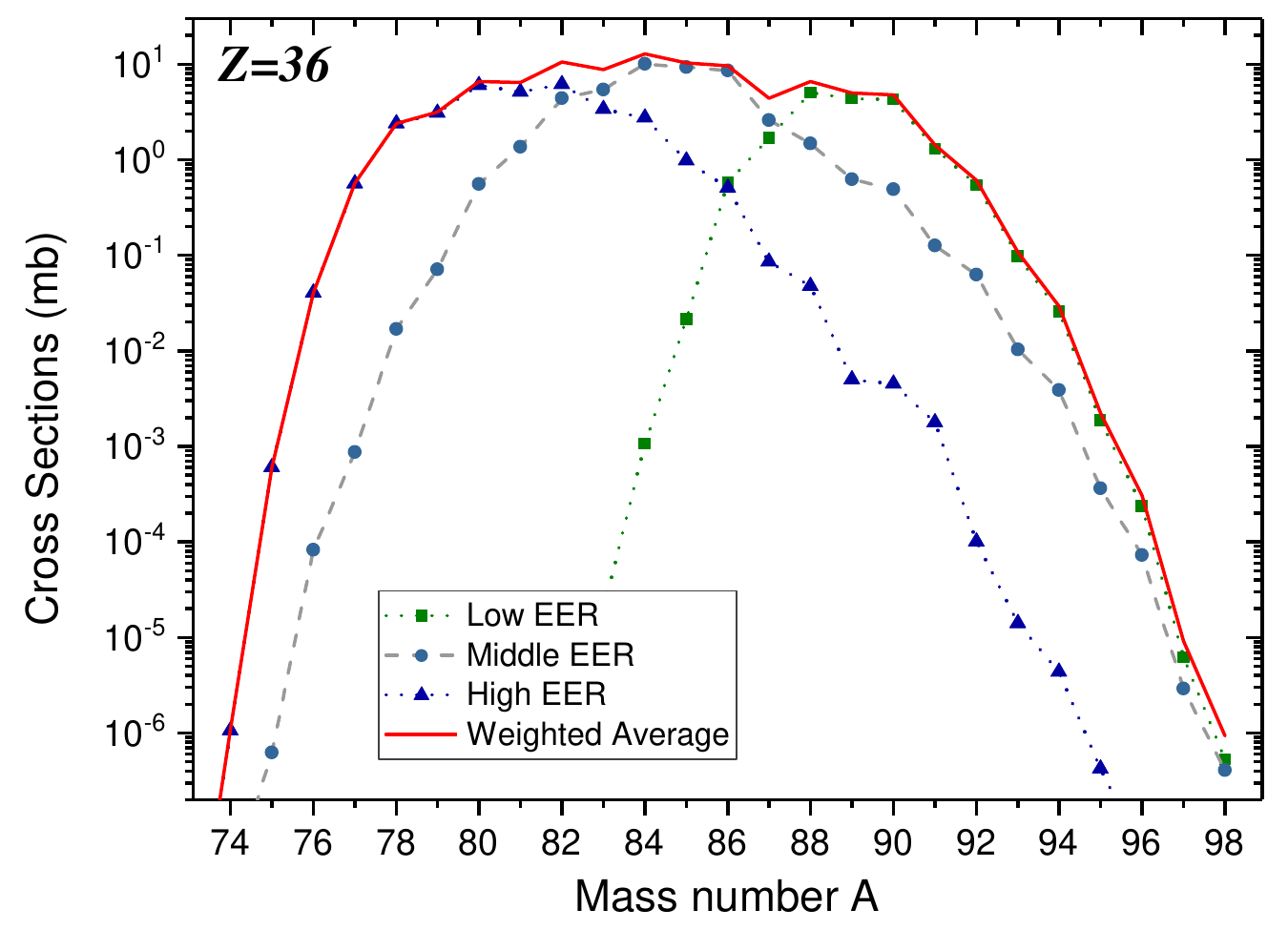}
\caption{The partial EER contributions to the production of krypton isotopes calculated with the \EER model for the abrasion step in the \nucl{238}{U} + \nucl{12}{C} reaction.}
\label{fig:partial_CS_Z36}
\end{figure}

\begin{figure}
\includegraphics[width=\columnwidth]{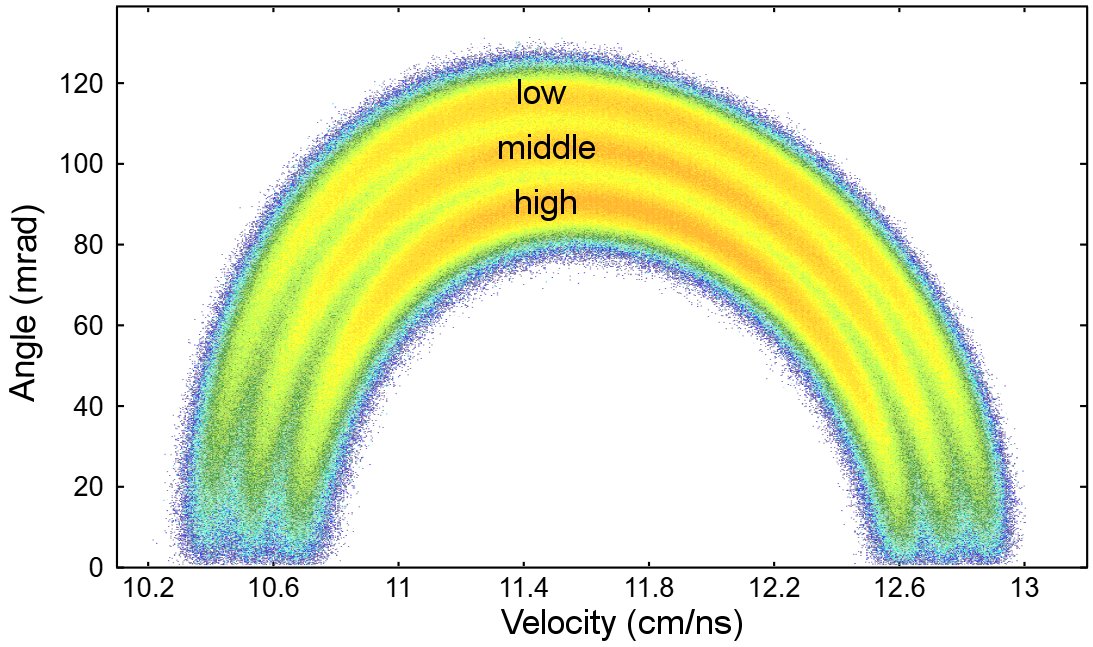}
\caption{\nucl{86}{Kr} kinematic distributions calculated for the different excitation energy regions defined in Table.~\ref{tab:eEER_param} for abrasion in the collision of a \nucl{238}{U} projectile interacting with carbon. In order to highlight the difference between each kinematic distribution, zero primary beam emittance and a thin target (0.1~mg/cm$^2$) were assumed in the calculation.}
\label{fig:86Kr_kinematics}
\end{figure}

Different parent nuclei can lead to the formation of the same final fission fragment, which 
complicates the experimental analysis. This is because the ion transmission needed to extract cross sections depends on the angular and momentum distributions of the fission fragment as determined by the properties of the parent nucleus ($Z, N, E_x$). This is illustrated in Figure~\ref{fig:partial_CS_Z36} which shows the partial EER contributions to the production of krypton isotopes. The kinematic distributions of $^{86}$Kr produced by different EER are displayed in Figure~\ref{fig:86Kr_kinematics}. Fragments with a fission progenitor belonging to the low EER and emitted in the forward (beam) direction possess the highest velocities. This is due to the enhanced Coulomb repulsion between fragments with a higher average $Z$ value.

The \lisepp \EER abrasion-fission model was previously used to interpret experimental data obtained at a high-energy fragmentation facility \cite{Suzuki13}. Here, it is used to analyze intermediate-energy abrasion-fission reactions on a thin $^{12}$C target, particularly with respect to the transmission and momentum space of the reaction products. To obtain cross sections, an average ion transmission weighted by each EER contribution is used. The production of fission fragments with different charge states must also be considered, where the formation of charge states is more likely compared with high-energy fragmentation, particularly at high $Z$.

\section{\label{level_exp}Experiment and analysis}

\begin{figure}[h]
\includegraphics[width=\columnwidth]{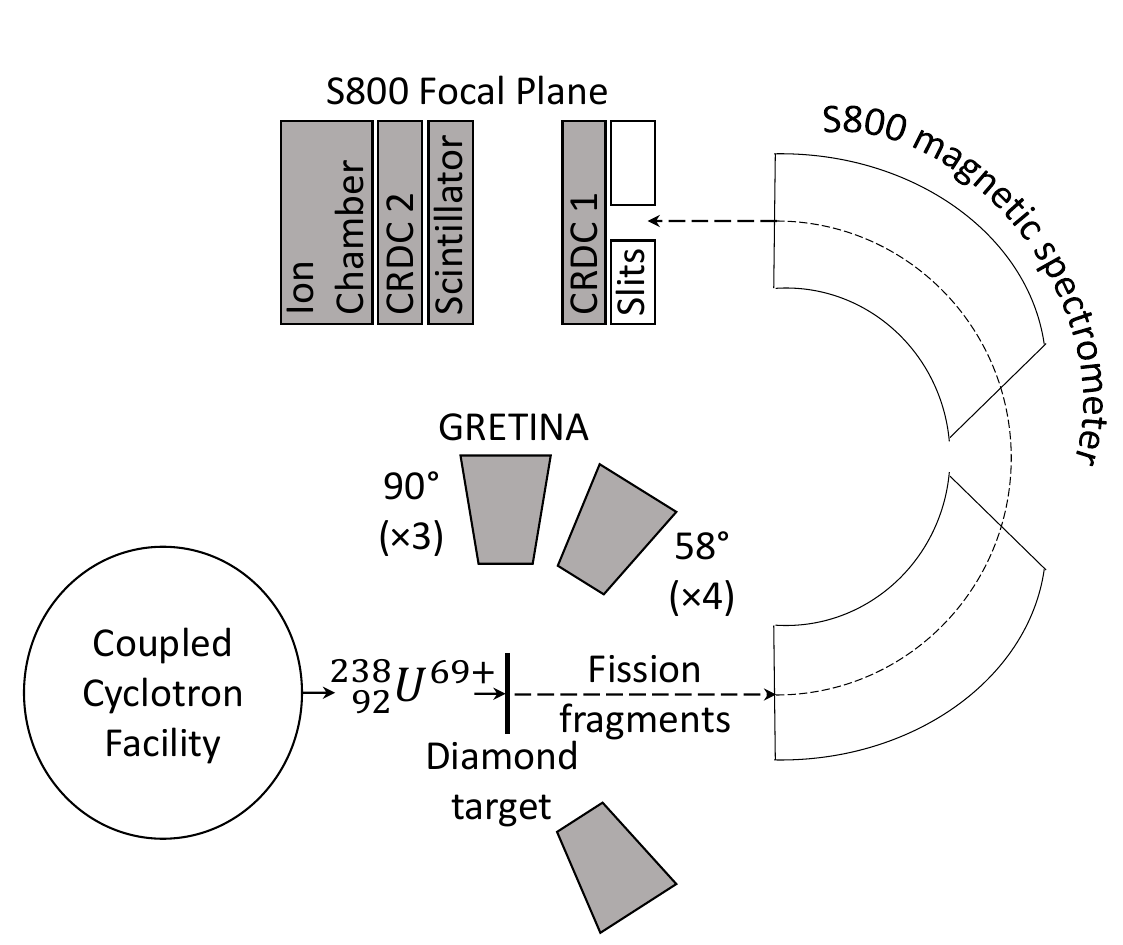}
\caption{Schematic showing the experimental setup used in the current work.}
\label{fig:experiment_setup}
\end{figure}

Fission fragments were produced using a \nucl{238}{U} beam provided by the Coupled Cyclotron Facility (CCF) at the National Superconducting Cyclotron Laboratory (NSCL) at Michigan State University. The \nucl{238}{U} beam was extracted from the K1200 cyclotron with $E$~=~80~MeV/u and charge state $q=69^{+}$. The beam was delivered directly to the \nucl{12}{C} reaction target located at the pivot point of the S800 magnetic spectrograph~\cite{Bazin03}. The target was constructed from a $20\times20$~mm$^2$ piece of electronic grade polycrystalline diamond, manufactured by Element Six (Cambridge, MA), with a nominal thickness of 100~$\mu$m (35~mg/cm$^{2}$). The experimental setup is shown in Figure~\ref{fig:experiment_setup}.\par

Gamma rays emitted by reaction residues following the de-population of excited nuclear states were detected using the Gamma-Ray Energy Tracking In-Beam Nuclear Array (GRETINA), providing both high efficiency (7.5~$\%$ at 1.3~MeV) and high energy resolution measurements for in-flight nuclear spectroscopy~\cite{Paschalis13}. GRETINA comprised 28 co-axial 36-fold electronically segmented high-purity germanium crystals arranged in seven modules around the target position, each module containing four crystals. A total of four out of seven GRETINA modules were placed at 58$\degree$ with respect to the primary beam axis and the remaining three were placed at 90$\degree$. The 3-dimensional $\gamma$-ray hit positions determined in GRETINA via digital pulse shape analysis techniques were used to reconstruct the $\gamma$-ray energy in the rest frame of the reaction residues.\par

Reaction residues were identified at the focal plane of the spectrograph via time-of-flight (ToF) and energy-loss measurements. A scintillator provided a start signal to a TAC for the ToF measurement. The scintillator was constructed at the NSCL using a $300\times150\times0.3$~mm$^3$ pressed sheet of UPS-923A (BC-400 equivalent) plastic, manufactured by Scintillation Technologies (Shirley, MA). The same signal was used to trigger the readout of all focal plane detectors and served as the master event trigger during the experiment. In addition to serving as the means of production for fission fragments, the diamond target also provided the stop signal for the ToF measurement. It was shown previously with stable beam that diamond detectors have a fast signal rise time ($<$~1~ns) and provide excellent timing resolution on the order of 10~ps~\cite{Stolz06,Michimasa13} depending on the experimental configuration. The trajectory of reaction residues in both the dispersive (orthogonal to the magnetic dipole field) and non-dispersive directions was measured using two position-sensitive cathode readout drift chambers (CRDC) placed either side of the scintillator. Energy-loss measurements were performed using a 16-channel segmented ion chamber filled with C$_{3}$F$_{8}$ gas operated between 340-400~Torr. Reaction residues exiting the ion chamber were stopped in material located at the rear of the focal plane.\par

Two different magnetic rigidity settings of the spectrograph were used, referred to as the \textit{low-rigidity} ($B\rho$~=~3.174~Tm) and \textit{high-rigidity} ($B\rho$~=~3.343~Tm) settings hereafter. The high-rigidity setting corresponds to a total beam-on-target time of approximately 39 hours, during which around $3.3\times10^{6}$ fission fragments were detected. The high-rigidity setting used a momentum acceptance of $\approx$~0.3~$\%$, reduced in order to limit the rate of the $^{238}$U$^{88, 89+}$ primary beam charge states incident upon the focal-plane detectors. The low-rigidity setting utilized close to the full acceptance of the spectrograph ($\approx$~2.4~$\%$). Around $2.1\times10^{5}$ fission fragments were detected in the low-rigidity setting which ran for 19 minutes.

\subsection{\label{level2_id}Identification of fission fragments}

\begin{figure*}
\includegraphics[width=\textwidth]{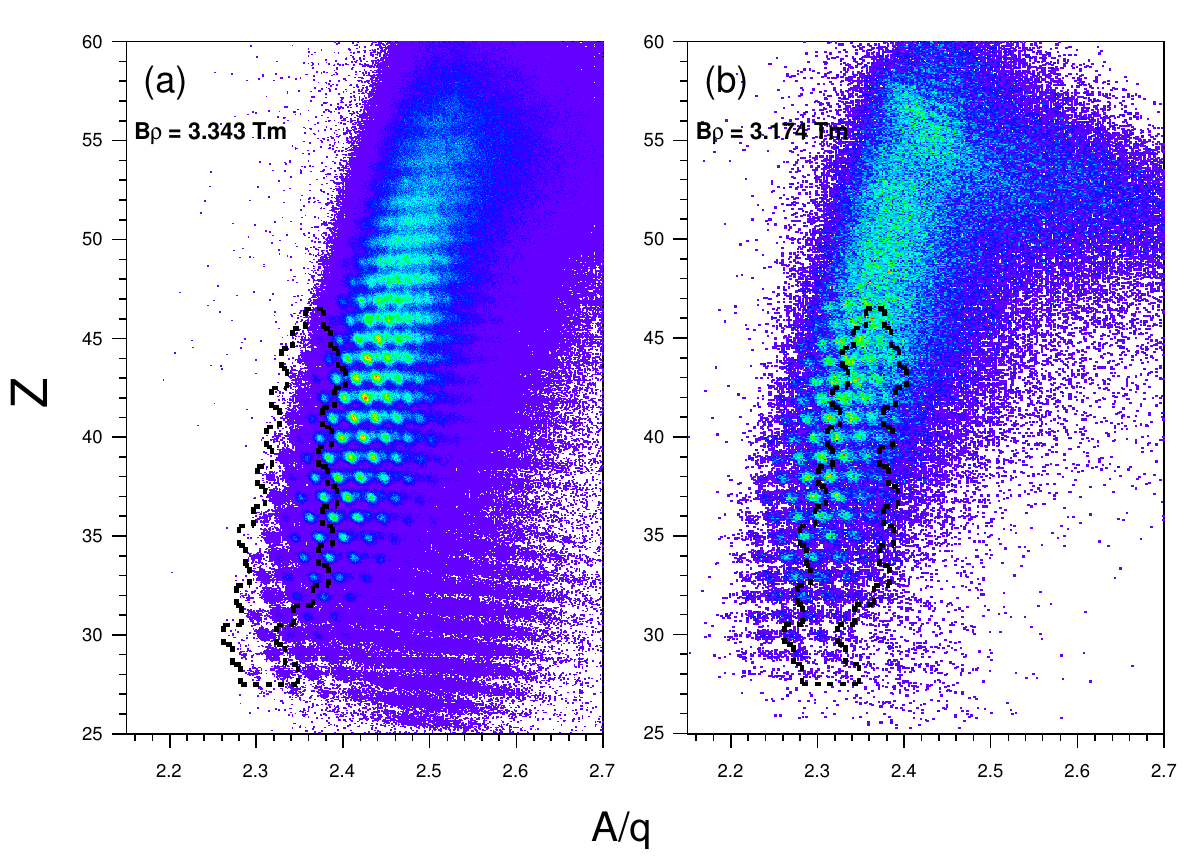}
\caption{Atomic number ($Z$) plotted against mass-to-charge ratio (\Aqsh) for fission fragments identified in the current work for the (a) high and (b) low magnetic rigidity settings of the S800 spectrograph. Fragments identified in both settings are shown enclosed by a dotted line. See text for details.}
\label{fig:pid}
\end{figure*}

Fission fragments identified in the current work are displayed in Figure~\ref{fig:pid} according to their atomic number $Z$ and mass-to-charge ratio \Aqsh. A total of 235 isotopes have been identified in both magnetic rigidity settings.
 $Z$ is corrected for the $(Z/\upsilon)^{2}$  dependence of the energy loss upon the ion velocity.
 An additional condition was imposed upon the energy loss registered by each CRDC anode to suppress ions that undergo changes in charge state in the focal-plane detector material. 
No reduction in efficiency for the detection of fission fragments in the ion chamber was observed upon requiring hits in both CDRCs.

The velocity of fission fragments was derived from ToF measurements between the diamond target and S800 focal plane scintillator using position and angle corrections obtained  with the CRDCs.
The path length traversed by ions following a central trajectory through the spectrometer was 14.3676~m.  

\begin{figure}[htp]
\centering
\includegraphics[width=\columnwidth]{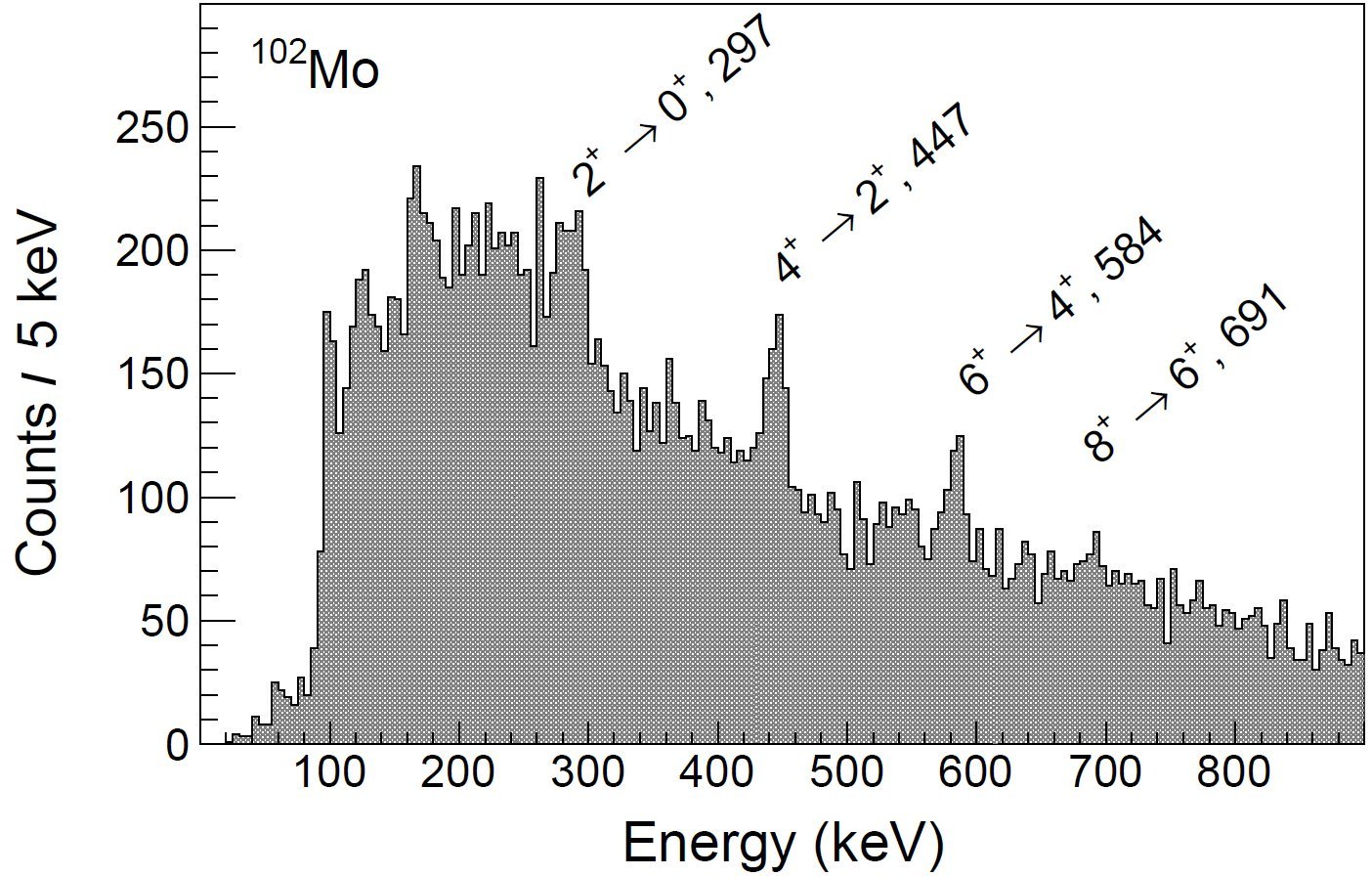}
\caption{Doppler-corrected $\gamma$-ray singles spectrum obtained for the fission fragment $^{102}$Mo identified in the high-rigidity setting.
 Data are restricted to GRETINA detectors located at 90$\degree$ relative to the beam direction. 
An additional condition has been imposed upon the interaction depth of the $\gamma$ rays in each HPGe crystal in order to suppress x rays.}
\label{fig:gamma}
\end{figure}

$Z$ and \Aq assignment in the high-rigidity setting proceeded via the correlation of ions detected 
at the focal plane with $\gamma$ rays observed in GRETINA. An example is shown in Figure~\ref{fig:gamma} where $\gamma$ ray transitions belonging to the ground-state (yrast) band in $^{102}$Mo are distinguished in the Doppler-corrected spectrum ($\upsilon/c\approx0.4$). The broad line width of the 297~keV transition may be due to the enhanced lifetime of the 2$^{+}$ state ($\tau=180$~ps) relative to higher-lying excited states in this nucleus. \par

No $\gamma$ rays were identified in the low-rigidity setting. In this case, $Z$ and $A/q$ assignments were cross-checked by comparing the experimental
production yields with those predicted by the \lisepp  software package  \cite{Tarasov08} (section \ref{level2_AF}). It is possible to determine experimental yields (and cross sections) independently in both settings. However, the identification of $\gamma$ rays in the high-rigidity setting directly assists our analysis of the low-rigidity setting.\par

Fission fragments that are fully-stripped of electrons provide the dominant contribution to the production yield below $Z\approx40$. A total kinetic energy measurement of the fragments was not possible in the current experiment, meaning the fully-stripped and hydrogen-like (+1e$^{-}$) charge states could not be separated on an event-by-event basis. However, the partial yields of each charge state can still be estimated via conventional means (see section \ref{level2_partial}). Particle identification beyond $Z\approx50$ was limited due to the increasing contribution of charge states to the production yield and the energy deposition in the ion chamber, where the latter begins to vary steeply with increasing depth for high-$Z$ fragments.\par

\subsection{\label{level2_partial}Extraction of partial ion yields}

\begin{figure}
\includegraphics[width=\columnwidth]{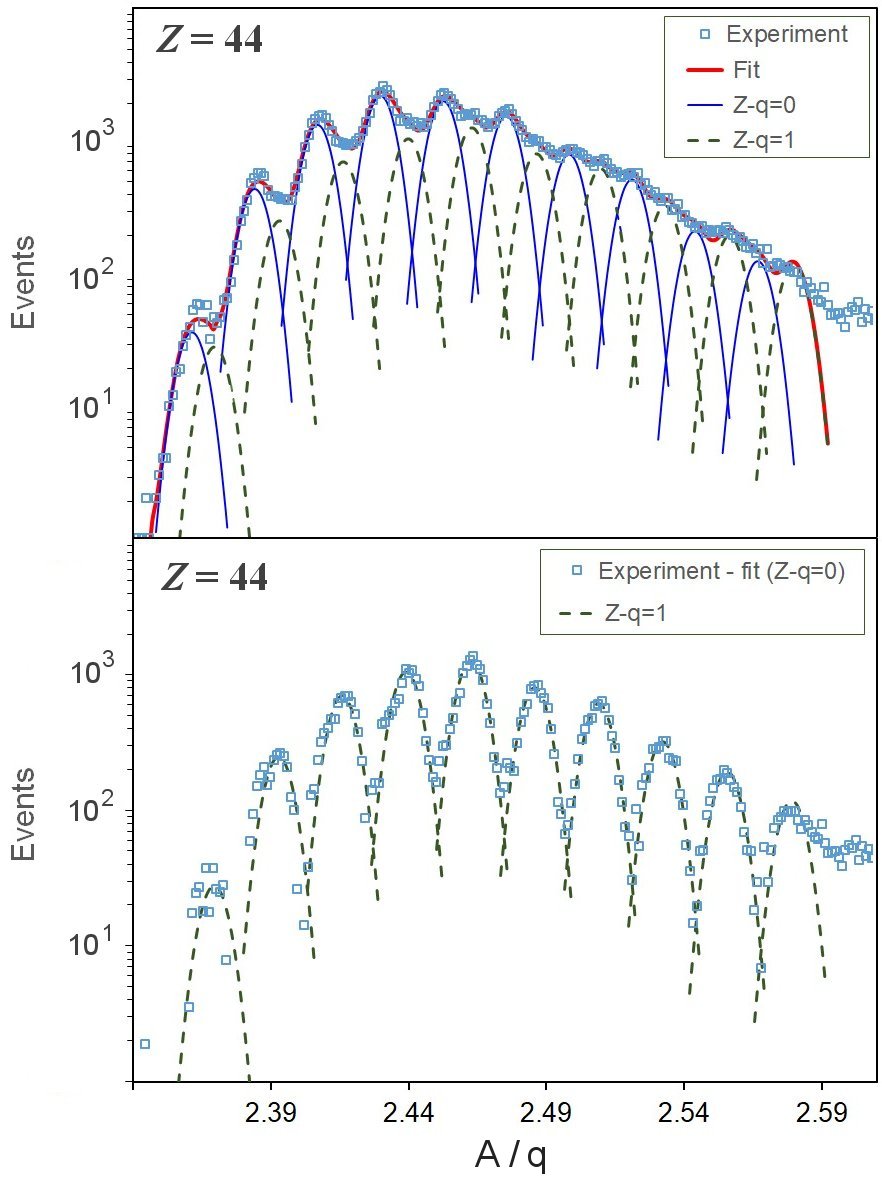}
\caption{Top: Partial yield analysis of ruthenium ions produced in the high-rigidity setting using a set of convoluted Gaussian functions. Peaks corresponding to fully-stripped ions (blue solid lines) 
and to hydrogen-like ions (green dashed lines) are included in the total fitted function (red solid line). Bottom: same as above, but only the Gaussian functions for the hydrogen-like ions and the experimental yield after subtraction of the fully-stripped contributions are shown. }
\label{fig:fitAq}
\end{figure}

\begin{figure}
\includegraphics[width=\columnwidth]{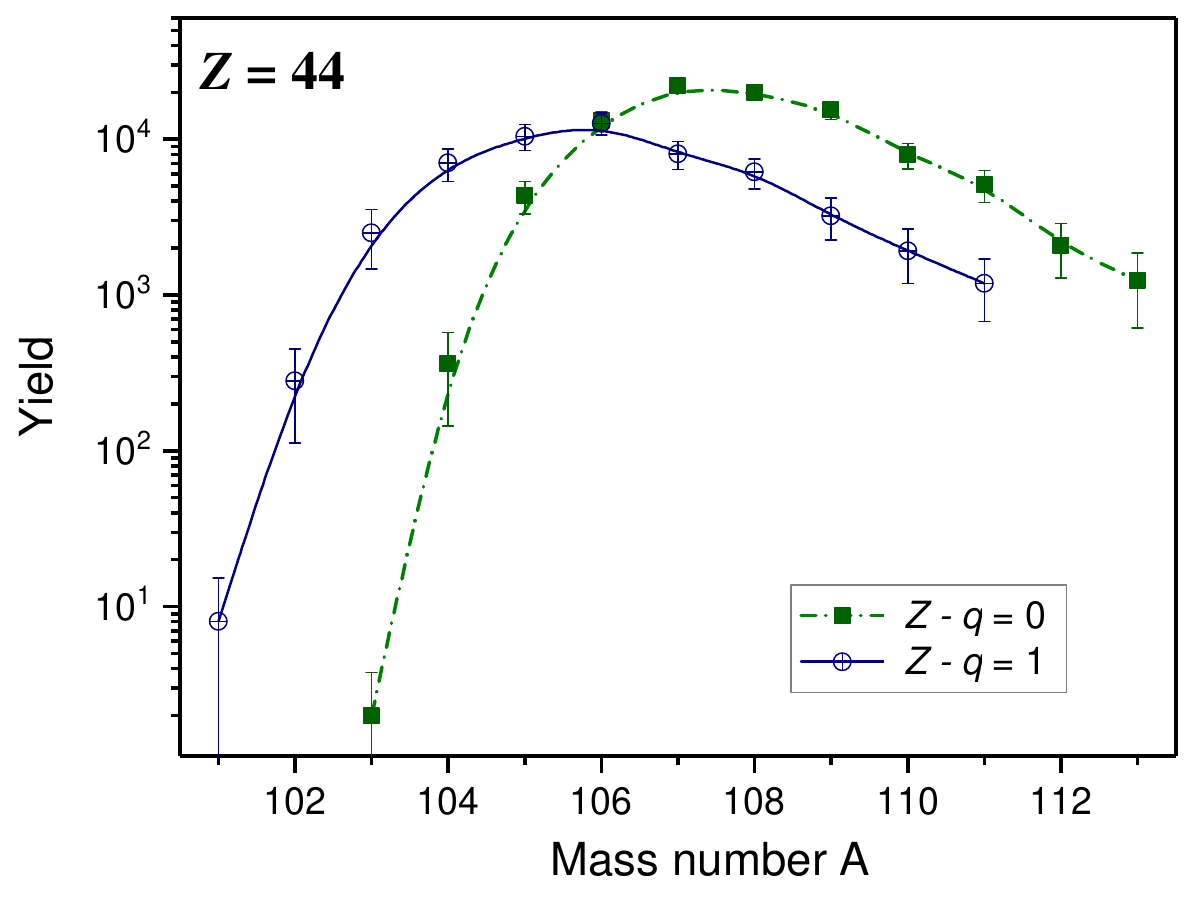}
\caption{ Ruthenium ion yields as resulting from the analysis shown in Figure~\ref{fig:fitAq}.}
\label{fig:yieldZ44}
\end{figure}

A minimization analysis for the extraction of ion yields from experimental spectra was applied due to the unavailability of a total kinetic energy measurement in this experiment and the moderate \Aq resolution ($\sigma$(\Aqsh)=0.0049). 
 The \Aq spectrum of each isotopic chain was isolated by gating on $Z$.  An example is shown in Figure~\ref{fig:fitAq} for the ruthenium isotopes. The hydrogen-like ions occupy interstitial positions between the higher-intensity fully-stripped products.
The \Aq spectra were fit with a spline comprised of several convoluted Gaussian functions in order to estimate the contributions of both the fully-stripped and hydrogen-like ions. 
Ion masses were used in the fitting process instead of integer mass numbers and improved the agreement of the calculated peak centroids with the experimental data.
The same width was used for all Gaussian ion functions. The fit allows for only one position parameter (representing a tiny $\delta$\Aq deviation due to calibration uncertainties). Thus, only $n+2$ parameters are varied to obtain the yields of $n$ ions with this approach.

Ruthenium ion yields are summarized in Figure~\ref{fig:yieldZ44}.

We note that, unfortunately, during the high-rigidity runs, the mylar stripper foil behind the gold electrode of the diamond target tore and changed the conditions for the fission fragments emerging from the target. However, the impact on the ion yields is minimal. The effect of the stripper foil is discussed in more detail in section \ref{level_charge}.


\subsection{\label{level2_yld}Fission fragment yield}

The fission fragment yield per incident beam particle was determined in each rigidity setting as:

\begin{equation} 
Y(A,Z,q) = \frac{N(A,Z,q)}{S_{f}N_{FP}f_{1}f_{2}f_{3}},
\label{eq:yield}
\end{equation}

\begin{figure}
\includegraphics[width=\columnwidth]{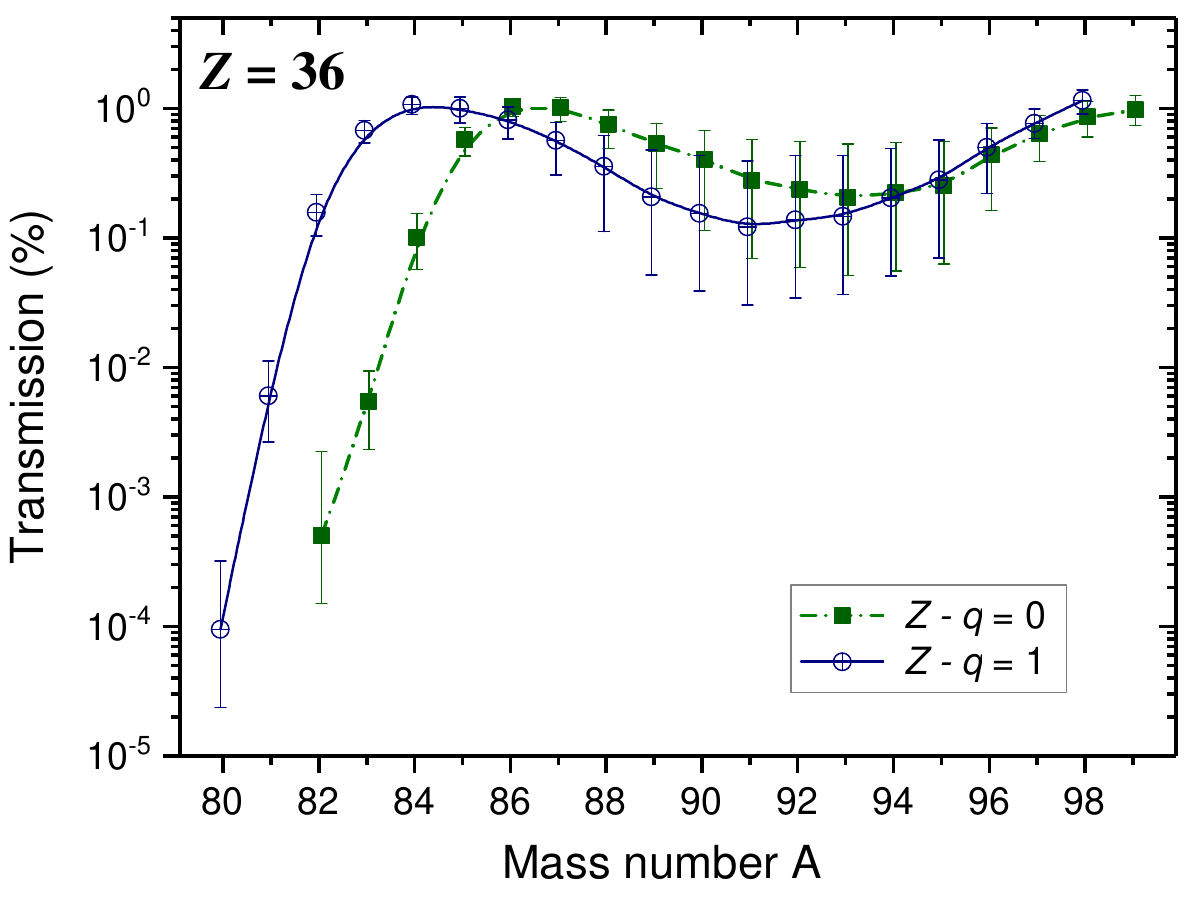}
\caption{Transmission ($\%$) of fully-stripped and hydrogen-like krypton fission fragments calculated with the \lisepp code. Charge-state factors were not taken into account.}
\label{fig:krtransmission}
\end{figure}

where $N(A,Z,q$) is the number of ions extracted for a given charge state from the fitted \Aq data (section \ref{level2_partial}), $N_{FP}$ is the total number of triggers registered by the focal-plane scintillator and $S_{f}$ is a scale factor relating the focal-plane triggers to the beam-on-target current. $N_{FP} = $\,6$\times$10$^{7}$ and 4.4\,$\times$10$^{5}$ for the high- and low-rigidity settings, respectively. The scale factor $S_f \approx$\,1900 in both settings. The correction factors $f_{1-3}$ are as follows: ratio of accepted to requested triggers at the focal plane ($f_1 \approx $\,0.9), fraction of incident uranium ions in true coincidence with focal-plane triggers ($f_2 \approx $\,0.8) and a $Z$-dependent focal-plane scintillator efficiency ($f_3$). The correction factor $f_3$ varied from 0.3 to 0.7 for $Z = 28 \rightarrow 34$, and was found to be unity for $Z\geq35$. The fission fragment yield is related to the cross section as:

\begin{equation} 
\sigma(A,Z,q) = \frac{Y(A,Z,q)}{n_{t}~~\varepsilon^*(A,Z,q)~~\psi(Z,q)}\:\times10^{27}\,[mb],
\label{eq:cross_sec}
\end{equation}

where $\varepsilon^*(A,Z,q)$ is the ion transmission through the spectrometer, $\psi(Z,q)$ is the charge-state production probability, and $n_{t}$ is the number of target atoms per square centimeter (1.68$\times$10$^{21}$\,cm$^{-2}$).
While the yield is wholly determined from experimental measurements, the ion transmission is not measured. The transmission must be calculated based on several factors including the mass ($A$), atomic number ($Z$), and charge ($q$) of the ion, the ion velocity and the ion-steering and focusing elements in the spectrometer. The ion momentum depends on the reaction mechanism, beam energy and energy-loss at the target position. Interactions at the target also affect the production of charge states and the likelihood of observing a residual nucleus with a given $A$, $Z$ at the focal plane depends on the relative contribution of each charge state. The transmission calculations are therefore model-dependent.
The ion transmission (deduced over all reactions) has been calculated with the \lisepp  code, and used in Equation~(\ref{eq:cross_sec}) without taking into account charge state 
factors~\footnote{
\begin{footnotesize}For this purpose, the \lisepp code was modified to provide a transmission deduced over all reactions for each charge state
 (Menu:  Files $\rightarrow$ Results $\rightarrow$ Transmission A,Z,q1 (summarized by reaction) without taking into account charge-state factors
\end{footnotesize}
}. 
Here and throughout the text, the superscript asterisk denotes that charge-state factors were not used to obtain this quantity.
The ion transmission $\varepsilon^*(A,Z,q)$ is calculated for a single charge state.
The calculated ion transmission for krypton isotopes is shown in Figure~\ref{fig:krtransmission} for the high-rigidity setting.
The transmission calculation details and transmission uncertainty analyses are discussed in section \ref{level3_err_trans}. The analysis of experimental charge-state factors and fission-fragment cross sections are discussed in sections \ref{level_charge} and \ref{level_cross_section}, respectively. 

\subsubsection{\label{level3_err_trans}Transmission uncertainties}

For the case of projectile fragmentation reactions, transmission calculations and transmission 
uncertainty analyses using the \lisepp code were discussed in Ref.~\cite{OT-NIMA09}.
For the present case of abrasion fission, the following variations were applied for the longitudinal selection transmission: target thickness and the `$f$' parameter, which determines the amount of excitation energy
taken out of the available fission Q-value~\cite{Faust02}.

\begin{table}
\setlength{\extrarowheight}{1.5pt}
\begin{center}
 \caption{ List of varied parameters used for transmission calculations and the systematic transmission error estimation.}
\label{Tab_LISE_runs}
\begin{footnotesize}
\begin{tabular}{|c|c|c|c|}
\hline
  Parameter &       Basic                   &    Minimum &   Maximum \\
                  &      configuration        &    set          &    set \\
\hline
\multicolumn{1}{|c|}{Angular  (H \& V)} 		   & \multirow{2}{*}{60 \& 100} & \multirow{2}{*}{50 \& 90} & \multirow{2}{*}{70 \& 100}  \\
\multicolumn{1}{|c|}{acceptance ($\pm$mrad) }  &                                           &                                          &  \\ 

\hline
High B$\rho$ FP slits (mm)  &     -6 : +22    &    -5 : +21      &   -7 : +23   \\
Low B$\rho$ FP slits (mm)  &     -116 : +140 &    -114 : +138      &   -118 : +142   \\
\hline
\multicolumn{1}{|c|}{Target thickness} 		     & \multirow{2}{*}{33.5} & \multirow{2}{*}{31.5} & \multirow{2}{*}{36}  \\
\multicolumn{1}{|c|}{ (mg/cm$^2$)  } &                                  &   &  \\ 
\hline
\multicolumn{1}{|c|}{`$f$' parameter} 		     & \multirow{2}{*}{4.5} & \multirow{2}{*}{3.5} & \multirow{2}{*}{5.5}  \\
\multicolumn{1}{|c|}{(fission Q-value~\cite{Faust02})  } &                                  &   &  \\ 
\hline
\end{tabular}
\end{footnotesize}
\end{center}
\end{table}

In order to estimate the systematic errors in the transmission
corrections, the angular and longitudinal selection
transmissions were computed with different parameters~(see
Table~\ref{Tab_LISE_runs}) for each isotope in both experimental
settings. 
The first \lisepp calculation with the basic configuration was
used to estimate the total fragment transmission including losses
due to reactions of the fragment in the target. Then, by varying one of the parameters as shown in Table~\ref{Tab_LISE_runs}, transmission uncertainties were calculated. A total of 9 calculations were performed (i.e. basic calculation plus 4 variations for the minimum and maximum settings) for each rigidity to determine the transmission uncertainties.

\begin{figure}
\centering
\includegraphics[width=\columnwidth]{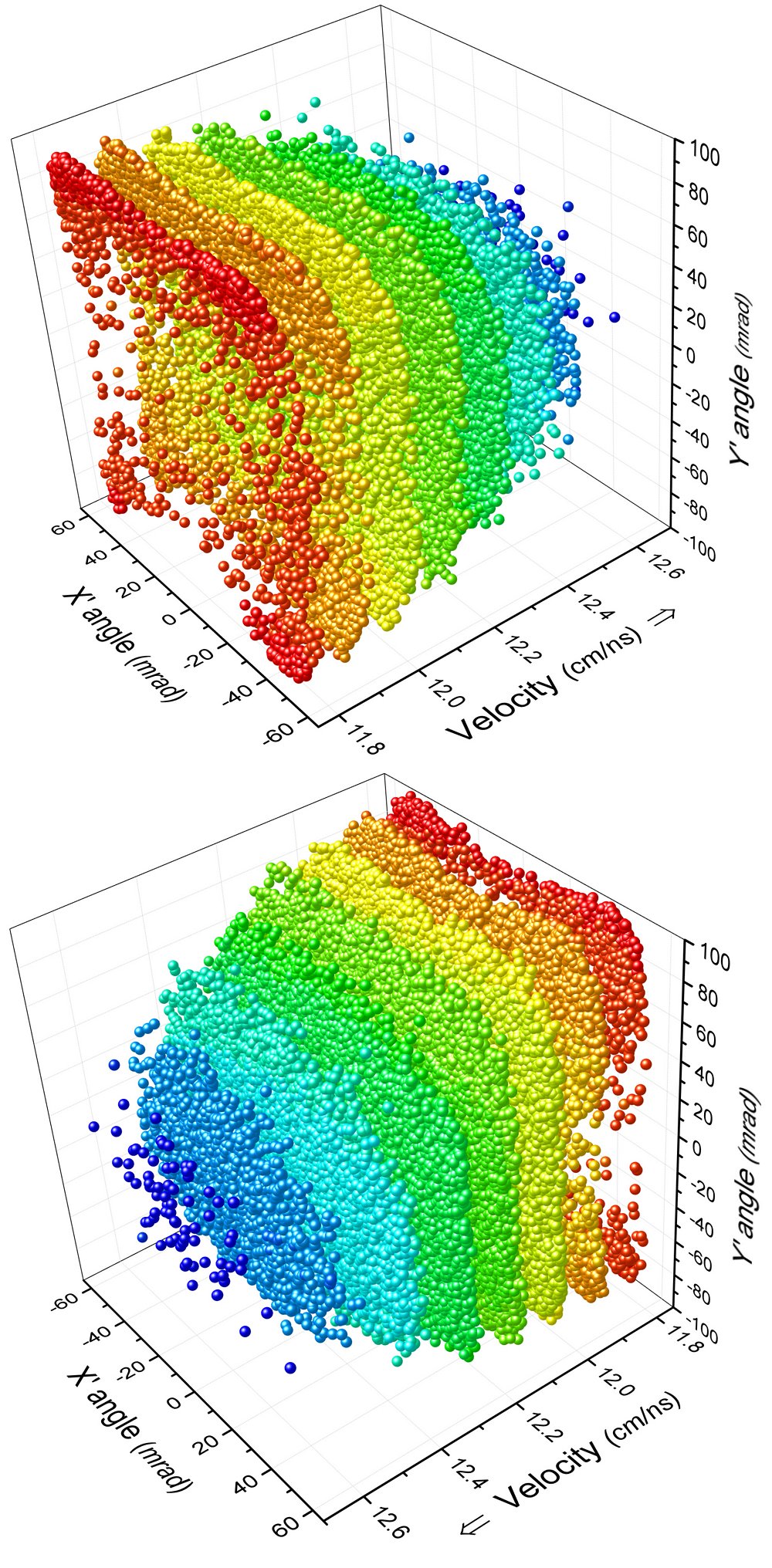}
\caption{Momentum space of krypton isotopes at the target position ($A$=83-90) reconstructed from experimental data using the \lisepp  software package. The reconstruction uses measurements of the velocity (ToF) and trajectory of the fission fragments performed at the focal-plane of the S800 spectrograph. Lighter masses appear at higher velocities. Data from the high-rigidity setting are shown. The top and bottom plots differ by a direction of the velocity axis.}
\label{fig:LISEreverse}
\end{figure}

\subsection{\label{level2_trj}Reconstruction of fission fragment trajectory}

As in spontaneous fission, nuclei induced to fission via the abrasion of one or more nucleons decay via binary fission producing two fission fragments. In the center-of-mass frame, each fragment is emitted at 180$\degree$ with respect to the other. Summing over all possible orientations of this decay vector leads naturally to the population of a spherical momentum shell \cite{Bernas03}, the thickness being determined by both the fission process and energy-loss in the target. In the laboratory frame, the shell appears elliptical and the fission fragments travel in the same direction as the projectile beam. Fragments emitted at backward angles possess a lower velocity relative to those emitted at forward angles.\par

\begin{figure}
\includegraphics[width=\columnwidth]{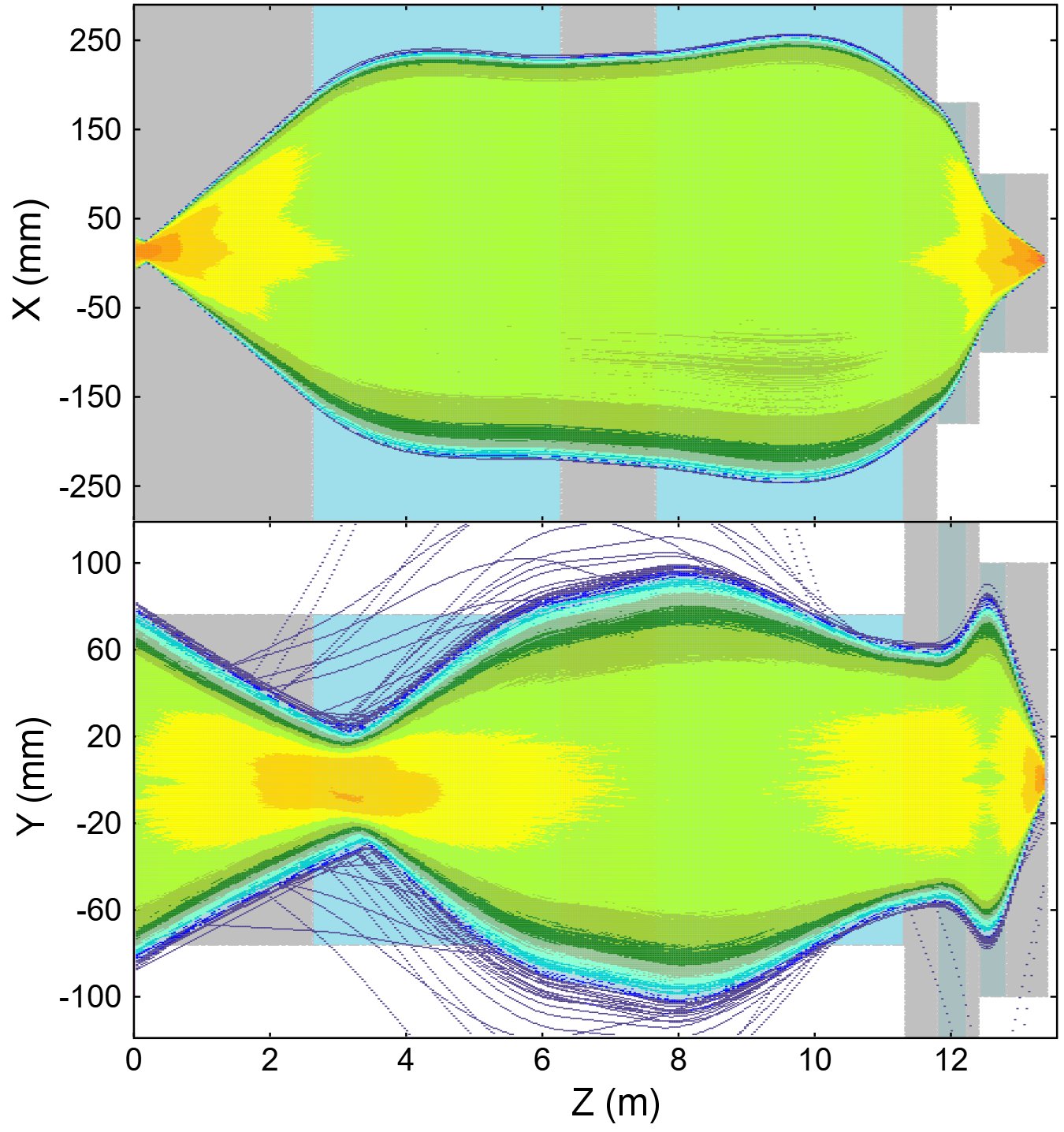}
\caption{The envelopes in the dispersive (top) and non-dispersive (bottom) planes produced with the reverse ray-tracing technique for fission products characterized in S800 focal plane detectors.  
The S800 focal plane corresponds to the z axis origin and the target is located at the end of the z axis. The blue and grey areas signify aperture sizes of the multipole and drift elements, respectively.}
\label{fig:s800_envelopes}
\end{figure}

\lisepp  is typically operated in so-called \textit{forward} mode where radioactive ions generated at the target position pass through various optical elements and/or detectors before stopping at the focal plane of the virtual spectrometer. In the current work, \lisepp  has been updated to  operate also in \textit{reverse} mode where experimental data is input to the simulation and the properties of reaction products extrapolated back toward the target position~\cite{LISE_Reverse_doc}. This capability allowed the direct comparison of the momentum space calculated in the abrasion-fission model with that extrapolated from data using the same software.
Velocity data were combined with measurements of the fission fragment trajectory at the focal plane in order to visualize the momentum space of the binomial reaction at the target position. Experimental data extracted from a \soft{ROOT} tree were written to a text file and input to the \lisepp  software package. The data were then transformed using the ion-optical parameters of the spectrometer read from a 5$^{th}$ order \soft{COSY} map~\cite{Makino06}. A reconstructed 3-dimensional image is shown in Figure~\ref{fig:LISEreverse} for krypton isotopes where the fragment velocity is plotted against the dispersive and non-dispersive angles $X^{'}$ and $Y^{'}$ in milliradians. Due to the narrow momentum acceptance used in the high-rigidity setting, each mass is represented by a distinct momentum cut. Lighter masses are observed at the most forward angles with the highest velocities (in blue) while heavier masses appear slower (orange-red) and begin to approach backward angles. It should be noted that while the dispersive angle was restricted during the high-rigidity setting, the non-dispersive angle had no such restriction and essentially used the full acceptance in this plane. Consequently, the $Y^{'}$ distribution is able to provide useful complementary information regarding the velocity of the fission fragments.\par

The reverse ray-tracing technique provides valuable benchmarks of the analysis providing the beam-optics constraints of fragments passing through a spectrometer.
Reverse rays should be inside of beam optics elements as can be seen from the reverse envelopes in the dispersive and non-dispersive planes plotted in Figure~\ref{fig:s800_envelopes}.
These envelopes demonstrate how the rays fit into the apertures of two quadrupole doublets located behind the target.

Fission fragment angle distributions obtained with the reverse ray-tracing technique will be used in a future analysis to deduce the parent nucleus velocity and then its mass and atomic number.


\section{\label{level1_rslt}Results and Discussion}

\subsection{\label{level_charge}Charge states}

At relativistic beam energies, fragments produced in nuclear reactions of projectile nuclei with target atoms emerge from the target mostly as fully-stripped ions. At lower beam energies, however, the probability to have electrons in the fragment's atomic orbitals increases, complicating the particle identification and requiring that charge-state factors have to be taken into account for cross-section analyses. The widely known charge-state evolution codes \softsh{GLOBAL}~\cite{GLOBAL}, \softsh{ETACHA4-GUI}~\cite{ETACHA4,LISE2023} and Winger's parameterization model~\cite{Intensity} are implemented 
in the \lisepp package and are used to calculate charge-state distributions and transmission estimations at intermediate energies.

The final fragment production cross section ($\sigma_F(A,Z)$) was calculated as the weighed average of the fragment production cross sections ($\sigma(A,Z,q)$)  obtained with  the different charge states. Based on Equation~\ref{eq:cross_sec}, each fragment production cross section can be considered as the ratio of the cross section ($\sigma^*_{i}$) for the formation of an ion without taking into account the charge-state production probability and its charge-state production probability ($\psi_{i}$):

\begin{equation} 
\sigma_{i} = \sigma^*_{i} ~ / ~ \psi_{i},
\label{eq:cs_charge1}
\end{equation}

Here, the index $i$ represents the charge state $Z-q$. Assuming that the fragment formation is independent of the charge state, which is equivalent to  $\sigma_{F} = \sigma_{1} = \sigma_{2}$, it is possible to determine the charge state factors from comparison of the production cross sections of fragment with different charge states. For this, it is important to impose the constraint that the charge-state factors should sum to one, ensuring proper normalization of the charge-state formation probabilities.

\begin{table}[h]
\caption{Calculations performed with the \soft{GLOBAL} code of the equilibrium charge state probabilities for Zn and Ru projectiles at 85 MeV/u after passing carbon and gold materials. The last column corresponds to the non-equilibrium case for fully-stripped Ru ions passing through a 0.15-micron thick gold foil. }
\label{Tab_ChargeDistr}
\setlength{\extrarowheight}{3pt}
\begin{center}
\begin{footnotesize}
\resizebox{0.95\columnwidth}{!}{
\begin{tabular}{|c|c|c|c|c|c|c|}
\hline

\multicolumn{2}{|l|}{Projectile} 	& Zn	& Zn	& Ru & Ru	& Ru$^{44+}$ \\
\multicolumn{2}{|l|}{Target}	& C	& Au	& C	& Au	& Au \\
\hline
\multicolumn{2}{|l|}{Equilibrium	}	&  \multicolumn{4}{c|}{yes} &	no \\
\hline
\multicolumn{2}{|l|}{Thickness, mg/cm$^2$}	&11.77&	1.05	&37.9	&2.83&	0.29\\
	    
\hline
	 &0	&97.8\%	&73.1\%	&82.1\%	&24.2\%	&58.7\% \\
\hspace{0.4cm} $Z-q$	\hspace{0.4cm} &1	&2.2\%	&24.8\%	&17.0\%	&46.5\%	&31.5\% \\
	&2	&0.01\%	&2.1\%	&0.9\%	&27.5\%	&8.8\% \\
\hline
\end{tabular}
}
\end{footnotesize}
\end{center}
\end{table}

It was previously mentioned in section \ref{level2_partial} that a mylar foil intended to strip the fragments emerging from the diamond target and gold anode was torn during the experiment. Therefore, charge-exchange processes in the gold anode must also be taken into account even if the foil was too thin to reach equilibrium in the charge state evolution. The charge-state probabilities for zinc and ruthenium projectiles at an energy of 85 MeV/u were calculated with the \soft{GLOBAL} code for carbon and gold foils and 
are listed in Table~\ref{Tab_ChargeDistr}. Helium-like ions are created with a probability of less than 10\%  after passing through a 150-nm thick gold foil. If, following the reaction, the fission fragments are mostly fully-stripped, and the typical metal layer thickness of a diamond detector contact is in the range of 30 to 200~nm, one may assume that $\psi_{0} + \psi_{1} \approxeq 1$. With this assumption, one can obtain charge state factors and then deduce a fission fragment cross section using experimental ion cross sections obtained using the  transmission without charge state factors  (see section~\ref{level2_yld}) with $\sigma_0 = \sigma_1$ in Equation~(\ref{eq:cs_charge1}), where 

\begin{equation} 
\begin{split}
&\sigma_0 =  \sigma^*_{0} ~/~  \psi_0 , \\
&\sigma_1 =  \sigma^*_{1} ~/~  \psi_{1} ~ \approxeq ~ \sigma^*_{1}  ~/~  (1-\psi_0).  
\end{split}
\end{equation}

The $\psi_0$ factors are deduced from matching the $\sigma_0$ and  $\sigma_1$ cross-section distributions for each element by minimizing the following sum:

\begin{equation}
S_Z =  \sum_{A_i}\:  \frac{ \left[\sigma_0 (A_i,Z)  - \sigma_1 (A_i,Z) \right]^2} {\sqrt[]{ \delta\sigma_0(A_i,Z)^2  + \delta\sigma_1(A_i,Z)^2 }} 
\label{eq:charge_minimization}
\end{equation}

\begin{figure}
\includegraphics[width=\columnwidth]{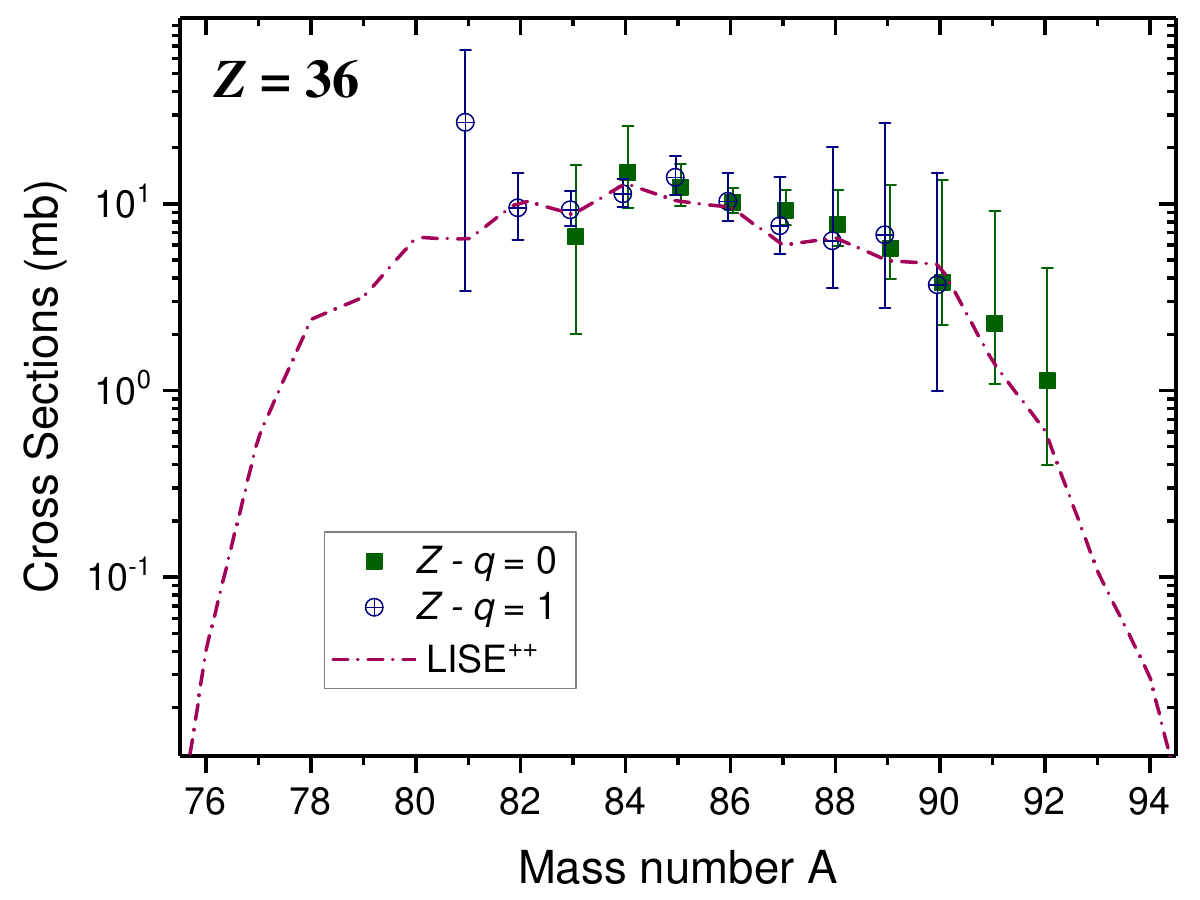}
\caption{Cross sections of krypton isotopes, fully-stripped and hydrogen-like, produced in the high-rigidity settings after matching with the charge-state factor $\psi_0 = 81.1^{+3.7}_{-4.5}$.
The cross sections calculated with \lisepp correspond to the weighted-average distributions in Figure~\ref{fig:partial_CS_Z36}.}
\label{fig:Brho1_Z36_matching}
\end{figure}

\begin{figure}
\includegraphics[width=\columnwidth]{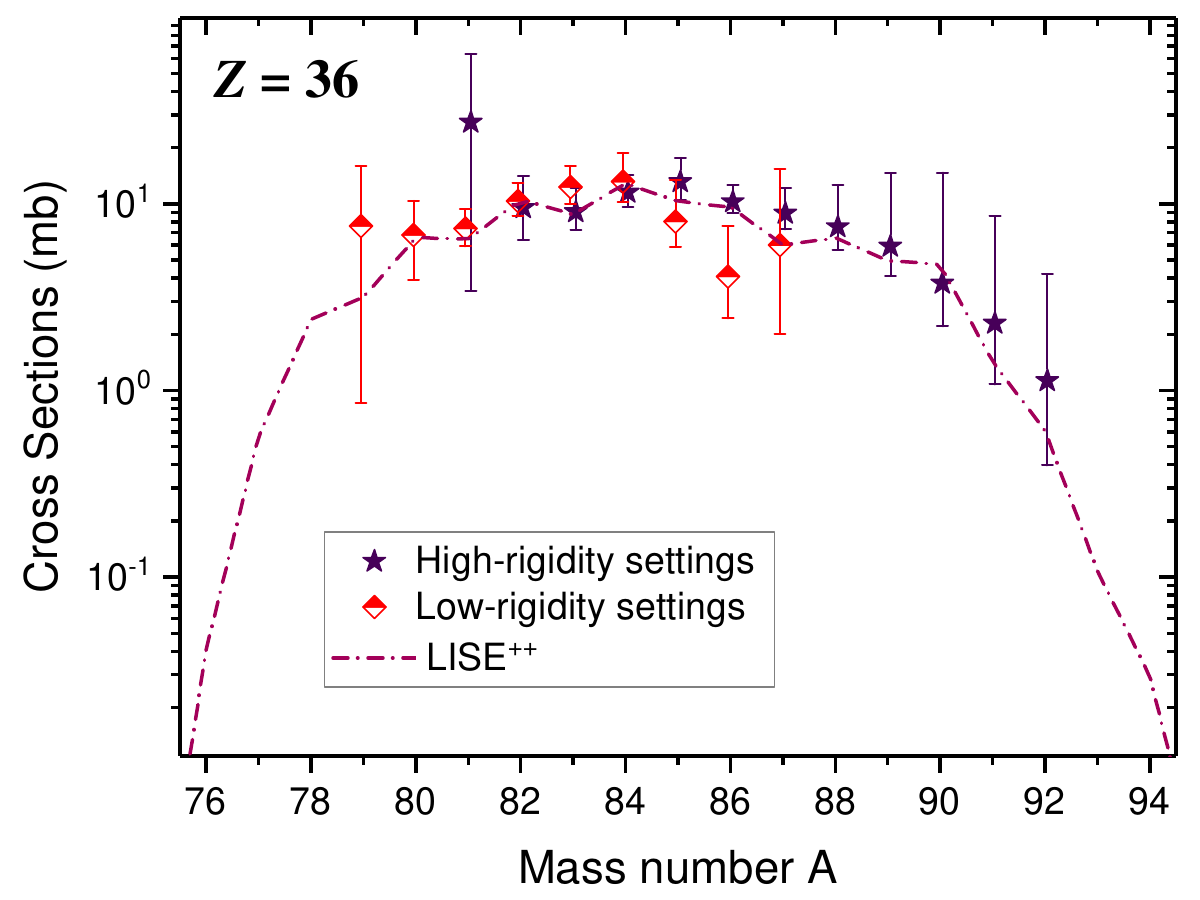}
\caption{Cross sections of krypton isotopes after the matching procedure (see Figure~\ref{fig:Brho1_Z36_matching}) produced in the high-rigidity and low-rigidity settings. 
The cross sections calculated with \lisepp correspond to the weighted-average distributions in Figure~\ref{fig:partial_CS_Z36}.}
\label{fig:Brho12_Z36}
\end{figure}

\begin{figure}
\includegraphics[width=\columnwidth]{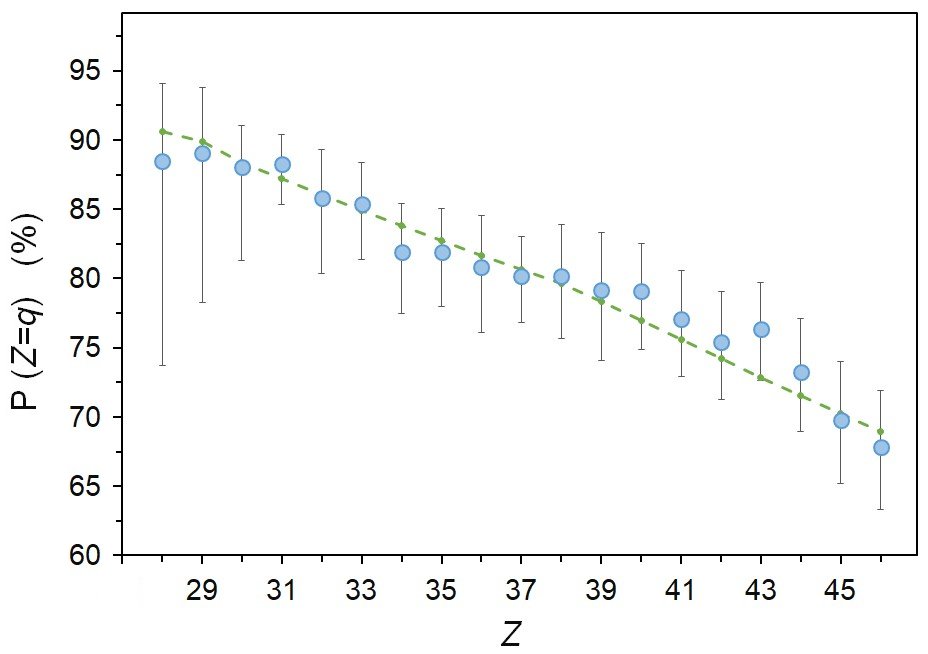}
\caption{Experimental elemental charge-state factors, $\psi_0$,  obtained from the high-rigidity settings. The \soft{GLOBAL} code charge-state calculations for passing though a gold foil with a thickness of 98~nm are shown as the green-dashed line.}
\label{fig:Brho1_ChargeStateFactors}
\end{figure}

An example of such a matching procedure is given in Figure~\ref{fig:Brho1_Z36_matching} for the krypton isotopes. Figure~\ref{fig:Brho12_Z36} shows the cross sections obtained following the matching procedure for isotopes produced in both rigidity settings.
The deduced elemental charge-state factors, $\Psi$, displayed in Figure~\ref{fig:Brho1_ChargeStateFactors}, 
show fairly good agreement with non-equilibrium charge-state calculations performed using the \soft{GLOBAL} code for passing a gold foil with a thickness of 98~nm. This value agrees with the expected range of the diamond detector contact thickness.

\begin{figure*}
\centering
\includegraphics[width=\linewidth]{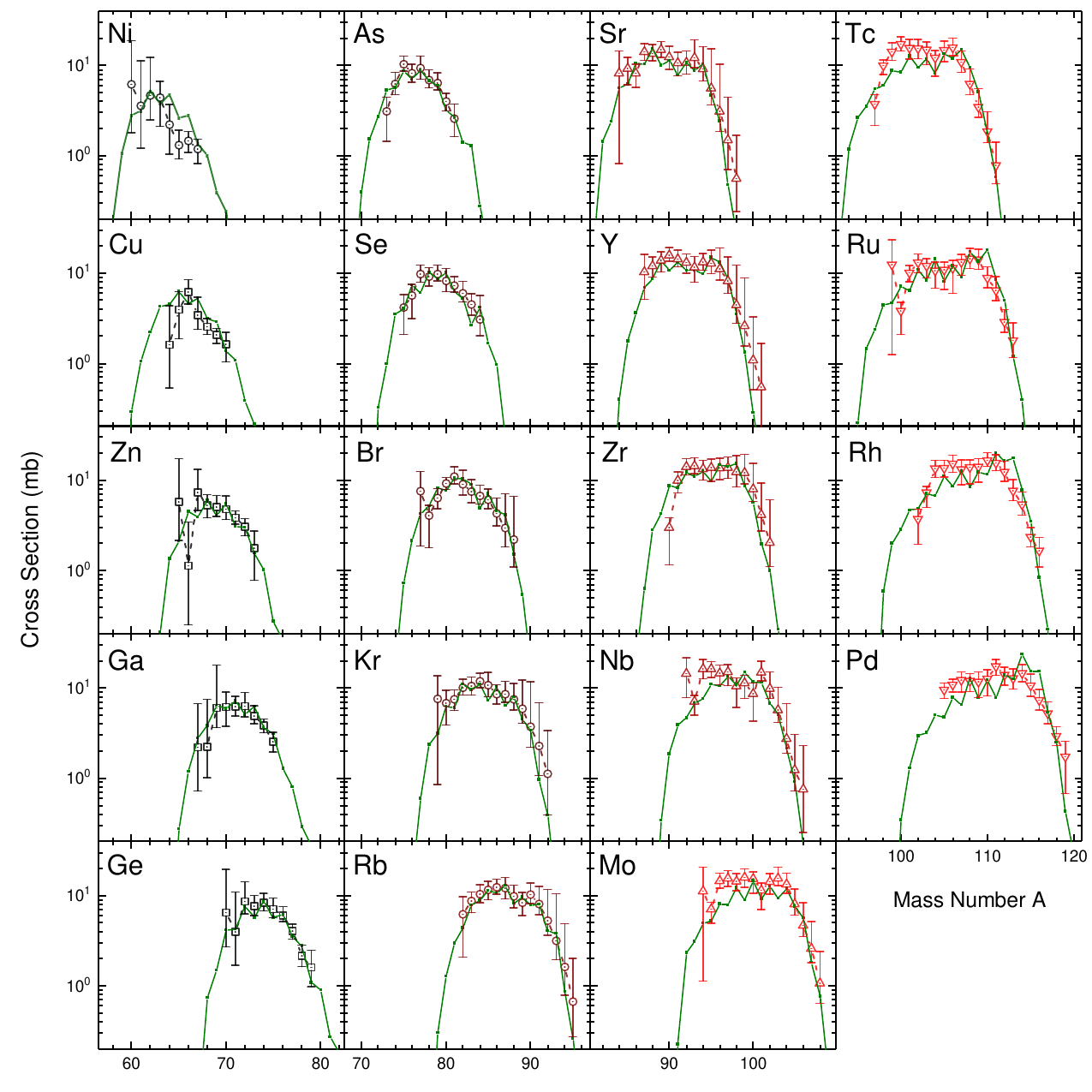}
\caption{Experimental isotopic cross sections of fission fragments for  nuclei identified in both  settings for 80 MeV/u uranium ions incident on a $^{12}$C target.
Cross-sections calculated with the \lisepp \EER model with parameters given in Table~\ref{tab:eEER_param} are shown as green solid lines.}
\label{fig:FinalCrossSections}
\end{figure*}


\subsection{\label{level_cross_section}Fission fragment cross sections}

\begin{figure}
\centering
\includegraphics[width=\columnwidth]{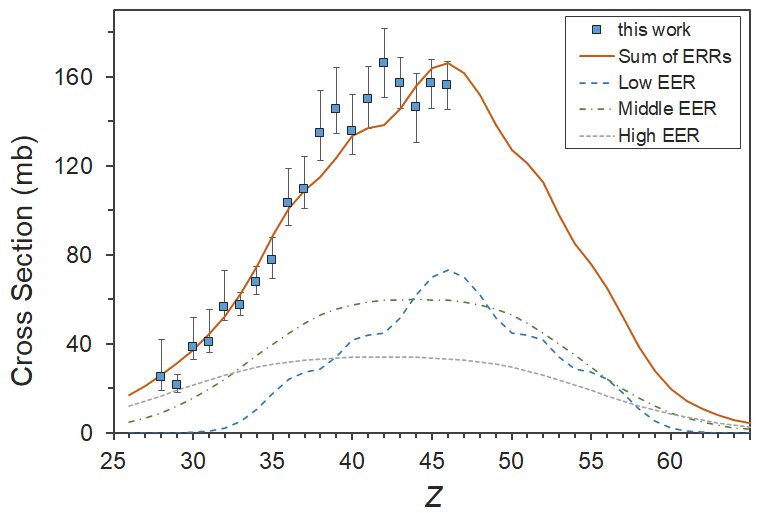}
\caption{Atomic number cross-section distributions of fission fragments produced by uranium at 80 MeV/u in this work and and as calculated with the \lisepp \EER model. 
}
\label{fig:CSbyZ}
\end{figure}

The isotopic production cross sections of fission fragments deduced from both rigidity settings are shown in Figure~\ref{fig:FinalCrossSections} for elements ranging from nickel $Z = 28$ to palladium $Z = 46$. The experimental cross sections are fairly well reproduced by 
calculations using the \lisepp \EER model, although some discrepancies are visible for high-$Z$ elements closer to stability which are produced by a highly-excited parent nucleus (see Figure~\ref{fig:partial_CS_Z36}).
It is worth noting that both the calculated cross sections and the experimental data depend on the 3EER model. In cases of poorer agreement where (for example) isotopes with $Z \approx$\,41-43 appear over-produced towards lower $A$ ($N$) compared to the calculated values, it is possible that the transmission was underestimated. While it is challenging to estimate transmission at the boundary of acceptance by the spectrometer, where it varies steeply as a function of rigidity, this effect can be characterised using the reverse-ray technique (section \ref{level2_trj}). Instead, the discrepancies may indicate a more complex set of contributions to the production of fission fragments than assumed by the current model.
Atomic-number cross-section distributions of fission fragments produced by uranium at 80 MeV/u in this work and as calculated by the \lisepp \EER model are shown in Figure~\ref{fig:CSbyZ}.
The shape of the total $Z$ distribution depends on the relative contributions of each parent nucleus. However, experimental sensitivity is reduced by the effect of averaging hundreds of parent nuclei in the EER model and representing them by a few progenitor nuclei: for example, \nucl{237}{U} with an excited energy of $41\,$MeV in the case of the low EER. To overcome this effect, a new Abrasion-Fission model taking into account contributions from all possible parent nuclei, is currently under development within the \lisepp framework.

\begin{figure}
\centering
\includegraphics[width=\columnwidth]{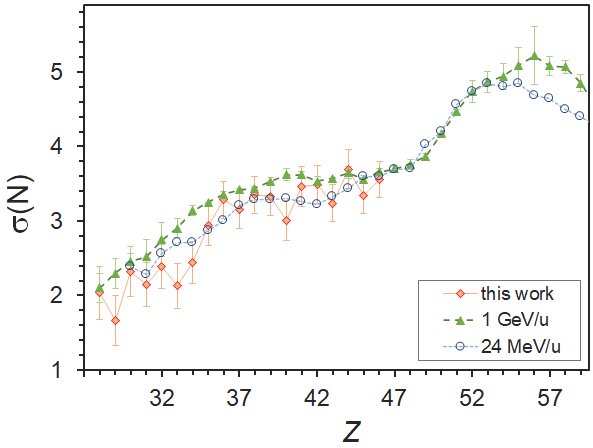}
\caption{Distributions of the neutron widths $\sigma_N$  compared with low- and high-primary-beam energy results with light targets~\cite{OT-EPJA18,Pereira07}}
\label{fig:sigmaN}
\end{figure}

The widths of the fission fragment distributions produced
by uranium in this work are shown as a function
of atomic number and are compared with the corresponding low- and high-energy results with light targets~\cite{OT-EPJA18,Pereira07} in Figure~\ref{fig:sigmaN}. The widths from the current work are found to have the same general trend with increasing $Z$ as the cited work.

\begin{figure}
\centering
\includegraphics[width=\columnwidth]{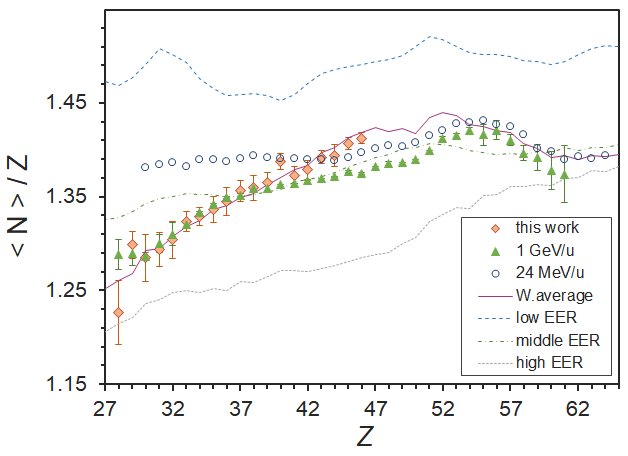}
\caption{Mean $\left<N(Z)\right>/Z$  ratios as a function of $Z$ for fission fragments produced by \nucl{238}{U} on  \nucl{12}{C} at 80 MeV/u (this work), with energy 24~MeV/u on C \cite{OT-EPJA18},  and with an energy 1~GeV/u on deuterium~\cite{Pereira07}.
The partial EER distributions and their weighted average results calculated with \lisepp \EER are indicated by the lines. }

\label{fig:NZ}
\end{figure}

The mean $\left<N(Z)\right>/Z$ ratios are shown in Figure~\ref{fig:NZ} and clearly indicate
that more neutron-rich isotopes of elements below
$Z = 42$ are produced with a carbon target at lower beam energy~\cite{OT-EPJA18}, whereas the results from the current experiment are close to the high-energy results~\cite{Pereira07}. 
To understand this observation, one first has to consider the high excitation energy of the fission fragments produced following abrasion-fission reactions which favors neutron evaporation and leads to a reduction of the $\left<N\right>/Z$ of the fragments. Other types of beam-target interactions, such as fast-fission, are expected to produce fission fragments at lower excitation energies and thus may enhance the proportion of neutron-rich fragments. Fast-fission involves the formation of a di-nuclear system with a vanishing fission barrier at small separation (impact parameter) of the projectile and target nuclei. Towards higher beam energies, the cross section for fast-fission is expected to decrease due to the high excitation energy of the residual nuclei, resulting in multi-fragmentation (or the `break-up’ channel within the abrasion-ablation
model). Fast-fission was previously demonstrated to be an important production mechanism for fragments below $Z \approx $~40 on carbon targets at 24~MeV/u bombarding energy \cite{OT-EPJA18}. This may indicate that there is a smaller contribution of the fast-fission component in the reactions of uranium at energies 80~MeV/u on a carbon target.
 
The mean $\left<N(Z)\right>/Z$ ratios resulting from the current work are fairly well reproduced by the \EER model (see `Weighted average' line in Figure~\ref{fig:NZ}), and the dominance of the high-excitation-energy yields over the low ones is apparent from the partial EER distributions up to $Z \approx 37$.

\section{\label{level4}Conclusions and outlook}

The properties of over 200 fission fragments have been measured following abrasion-fission reactions of a 80~MeV/u \nucl{238}{U} beam on a thin $^{12}$C (diamond) active target. Total yields obtained for nuclei in the range $Z=28-46$ and $A\approx 60-110$, approximately half of the expected mass distribution of reaction products, provide around 10$^{-5}$ to 10$^{-4}$ fission fragments per incident uranium ion depending on the rigidity and acceptance settings of the spectrometer. 
Partial yields indicate significant production of the hydrogen-like charge state relative to fully-stripped ions of 10-30$\%$. The separation of charge states (e.g. using total kinetic energy) should therefore be considered a vital experimental tool for the unambiguous identification of intermediate-energy fission fragments.\par

In particular the kinematic properties of fission fragments provide a sensitive probe of the initial conditions of the fissioning nucleus. An in-depth analysis of the laboratory-frame momentum space at the production target was performed using the \lisepp  software package, where the trajectory of an ion passing through the spectrometer can be reconstructed based upon measurements at the focal plane. A generally good agreement is observed between correlated velocity and angle measurements and those predicted by the \lisepp  \EER abrasion-fission model within uncertainty limits. 

Efforts are underway to develop an abrasion-fission model that includes a realistic ensemble of nuclei following peripheral collisions. Due to the large amount of nuclei involved in calculations, both speed and reliability will be important factors.

The experimental techniques demonstrated in the current work provide an ideal platform for future studies with the next generation of high-acceptance spectrometers such as the High Rigidity Spectrometer at FRIB~\cite{HRS}. These studies aim to simultaneously detect both fission products and obtain their velocities and angles at the reaction point. In particular, this will allow for the unique identification of the fissioning nucleus after the abrasion of the projectile.

\begin{acknowledgments}
The expertise of NSCL operations staff and of the A1900 separator group is gratefully acknowledged. The authors are thankful for the fruitful discussions held with Prof.~D.J.~Morrissey and Dr.~E.~Kwan during the analysis of this experiment. This work was supported by the US National Science Foundation under Grants No. PHY-15-65546 and No. PHY-20-12040 and by the U.S. Department of Energy, Office of Science, Office of Nuclear Physics. under Grant No. DE-SC0020451 (MSU). GRETINA was funded by the DOE, Office of Science.
\end{acknowledgments}

\section*{References}
\bibliographystyle{apsrev4-2}
\bibliography{bibliography}

\begin{thebibliography}{43}%
\makeatletter
\providecommand \@ifxundefined [1]{%
 \@ifx{#1\undefined}
}%
\providecommand \@ifnum [1]{%
 \ifnum #1\expandafter \@firstoftwo
 \else \expandafter \@secondoftwo
 \fi
}%
\providecommand \@ifx [1]{%
 \ifx #1\expandafter \@firstoftwo
 \else \expandafter \@secondoftwo
 \fi
}%
\providecommand \natexlab [1]{#1}%
\providecommand \enquote  [1]{``#1''}%
\providecommand \bibnamefont  [1]{#1}%
\providecommand \bibfnamefont [1]{#1}%
\providecommand \citenamefont [1]{#1}%
\providecommand \href@noop [0]{\@secondoftwo}%
\providecommand \href [0]{\begingroup \@sanitize@url \@href}%
\providecommand \@href[1]{\@@startlink{#1}\@@href}%
\providecommand \@@href[1]{\endgroup#1\@@endlink}%
\providecommand \@sanitize@url [0]{\catcode `\\12\catcode `\$12\catcode
  `\&12\catcode `\#12\catcode `\^12\catcode `\_12\catcode `\%12\relax}%
\providecommand \@@startlink[1]{}%
\providecommand \@@endlink[0]{}%
\providecommand \url  [0]{\begingroup\@sanitize@url \@url }%
\providecommand \@url [1]{\endgroup\@href {#1}{\urlprefix }}%
\providecommand \urlprefix  [0]{URL }%
\providecommand \Eprint [0]{\href }%
\providecommand \doibase [0]{https://doi.org/}%
\providecommand \selectlanguage [0]{\@gobble}%
\providecommand \bibinfo  [0]{\@secondoftwo}%
\providecommand \bibfield  [0]{\@secondoftwo}%
\providecommand \translation [1]{[#1]}%
\providecommand \BibitemOpen [0]{}%
\providecommand \bibitemStop [0]{}%
\providecommand \bibitemNoStop [0]{.\EOS\space}%
\providecommand \EOS [0]{\spacefactor3000\relax}%
\providecommand \BibitemShut  [1]{\csname bibitem#1\endcsname}%
\let\auto@bib@innerbib\@empty
\bibitem [{\citenamefont {Goriely}\ \emph {et~al.}(2013)\citenamefont {Goriely}
  \emph {et~al.}}]{Goriely13}%
  \BibitemOpen
  \bibfield  {author} {\bibinfo {author} {\bibfnamefont {S.}~\bibnamefont
  {Goriely}} \emph {et~al.},\ }\href@noop {} {\bibfield  {journal} {\bibinfo
  {journal} {Phys. Rev. Lett.}\ }\textbf {\bibinfo {volume} {111}},\ \bibinfo
  {pages} {242502} (\bibinfo {year} {2013})}\BibitemShut {NoStop}%
\bibitem [{\citenamefont {Algora}\ \emph {et~al.}(2010)\citenamefont {Algora}
  \emph {et~al.}}]{Algora10}%
  \BibitemOpen
  \bibfield  {author} {\bibinfo {author} {\bibfnamefont {A.}~\bibnamefont
  {Algora}} \emph {et~al.},\ }\href@noop {} {\bibfield  {journal} {\bibinfo
  {journal} {Phys. Rev. Lett.}\ }\textbf {\bibinfo {volume} {105}},\ \bibinfo
  {pages} {202501} (\bibinfo {year} {2010})}\BibitemShut {NoStop}%
\bibitem [{\citenamefont {Schmidt}\ \emph {et~al.}(1993)\citenamefont {Schmidt}
  \emph {et~al.}}]{Schmidt93}%
  \BibitemOpen
  \bibfield  {author} {\bibinfo {author} {\bibfnamefont {K.~H.}\ \bibnamefont
  {Schmidt}} \emph {et~al.},\ }\href@noop {} {\bibfield  {journal} {\bibinfo
  {journal} {Phys. Lett. B}\ }\textbf {\bibinfo {volume} {300}},\ \bibinfo
  {pages} {313} (\bibinfo {year} {1993})}\BibitemShut {NoStop}%
\bibitem [{\citenamefont {Bernas}\ \emph {et~al.}(1997)\citenamefont {Bernas}
  \emph {et~al.}}]{Bernas97}%
  \BibitemOpen
  \bibfield  {author} {\bibinfo {author} {\bibfnamefont {M.}~\bibnamefont
  {Bernas}} \emph {et~al.},\ }\href@noop {} {\bibfield  {journal} {\bibinfo
  {journal} {Phys. Lett. B}\ }\textbf {\bibinfo {volume} {415}},\ \bibinfo
  {pages} {111} (\bibinfo {year} {1997})}\BibitemShut {NoStop}%
\bibitem [{\citenamefont {Enqvist}\ \emph {et~al.}(1999)\citenamefont {Enqvist}
  \emph {et~al.}}]{Enqvist99}%
  \BibitemOpen
  \bibfield  {author} {\bibinfo {author} {\bibfnamefont {T.}~\bibnamefont
  {Enqvist}} \emph {et~al.},\ }\href@noop {} {\bibfield  {journal} {\bibinfo
  {journal} {Nucl. Phys. A}\ }\textbf {\bibinfo {volume} {658}},\ \bibinfo
  {pages} {47} (\bibinfo {year} {1999})}\BibitemShut {NoStop}%
\bibitem [{\citenamefont {Schmidt}\ \emph {et~al.}(2001)\citenamefont {Schmidt}
  \emph {et~al.}}]{Schmidt01}%
  \BibitemOpen
  \bibfield  {author} {\bibinfo {author} {\bibfnamefont {K.-H.}\ \bibnamefont
  {Schmidt}} \emph {et~al.},\ }\href@noop {} {\bibfield  {journal} {\bibinfo
  {journal} {Nucl. Phys. A}\ }\textbf {\bibinfo {volume} {693}},\ \bibinfo
  {pages} {169} (\bibinfo {year} {2001})}\BibitemShut {NoStop}%
\bibitem [{\citenamefont {Bernas}\ \emph {et~al.}(2003)\citenamefont {Bernas}
  \emph {et~al.}}]{Bernas03}%
  \BibitemOpen
  \bibfield  {author} {\bibinfo {author} {\bibfnamefont {M.}~\bibnamefont
  {Bernas}} \emph {et~al.},\ }\href@noop {} {\bibfield  {journal} {\bibinfo
  {journal} {Nucl. Phys. A}\ }\textbf {\bibinfo {volume} {725}},\ \bibinfo
  {pages} {213} (\bibinfo {year} {2003})}\BibitemShut {NoStop}%
\bibitem [{\citenamefont {Pereira}\ \emph {et~al.}(2007)\citenamefont {Pereira}
  \emph {et~al.}}]{Pereira07}%
  \BibitemOpen
  \bibfield  {author} {\bibinfo {author} {\bibfnamefont {J.}~\bibnamefont
  {Pereira}} \emph {et~al.},\ }\href@noop {} {\bibfield  {journal} {\bibinfo
  {journal} {Phys. Rev. C}\ }\textbf {\bibinfo {volume} {75}},\ \bibinfo
  {pages} {014602} (\bibinfo {year} {2007})}\BibitemShut {NoStop}%
\bibitem [{\citenamefont {S.~Cohen}\ and\ \citenamefont
  {Swiatecki}(1974)}]{Cohen74}%
  \BibitemOpen
  \bibfield  {author} {\bibinfo {author} {\bibfnamefont {F.~P.}\ \bibnamefont
  {S.~Cohen}}\ and\ \bibinfo {author} {\bibfnamefont {W.~J.}\ \bibnamefont
  {Swiatecki}},\ }\href@noop {} {\bibfield  {journal} {\bibinfo  {journal}
  {Annals of Physics}\ }\textbf {\bibinfo {volume} {82}},\ \bibinfo {pages}
  {557} (\bibinfo {year} {1974})}\BibitemShut {NoStop}%
\bibitem [{\citenamefont {Goldhaber}(1974)}]{Goldhaber74}%
  \BibitemOpen
  \bibfield  {author} {\bibinfo {author} {\bibfnamefont {A.~S.}\ \bibnamefont
  {Goldhaber}},\ }\href@noop {} {\bibfield  {journal} {\bibinfo  {journal}
  {Phys. Lett. B}\ }\textbf {\bibinfo {volume} {53}},\ \bibinfo {pages} {306}
  (\bibinfo {year} {1974})}\BibitemShut {NoStop}%
\bibitem [{\citenamefont {Tarasov}(2004)}]{OT-NPA04}%
  \BibitemOpen
  \bibfield  {author} {\bibinfo {author} {\bibfnamefont {O.}~\bibnamefont
  {Tarasov}},\ }\href@noop {} {\bibfield  {journal} {\bibinfo  {journal} {Nucl.
  Phys. A}\ }\textbf {\bibinfo {volume} {734}},\ \bibinfo {pages} {536}
  (\bibinfo {year} {2004})}\BibitemShut {NoStop}%
\bibitem [{\citenamefont {Ohnishi}\ \emph {et~al.}(2008)\citenamefont {Ohnishi}
  \emph {et~al.}}]{Ohnishi08}%
  \BibitemOpen
  \bibfield  {author} {\bibinfo {author} {\bibfnamefont {T.}~\bibnamefont
  {Ohnishi}} \emph {et~al.},\ }\href {https://doi.org/10.1143/JPSJ.77.083201}
  {\bibfield  {journal} {\bibinfo  {journal} {J. Phys. Soc. Jpn.}\ }\textbf
  {\bibinfo {volume} {77}},\ \bibinfo {pages} {083201} (\bibinfo {year}
  {2008})}\BibitemShut {NoStop}%
\bibitem [{\citenamefont {Ohnishi}\ \emph {et~al.}(2010)\citenamefont {Ohnishi}
  \emph {et~al.}}]{Ohnishi10}%
  \BibitemOpen
  \bibfield  {author} {\bibinfo {author} {\bibfnamefont {T.}~\bibnamefont
  {Ohnishi}} \emph {et~al.},\ }\href@noop {} {\bibfield  {journal} {\bibinfo
  {journal} {J. Phys. Soc. Jpn.}\ }\textbf {\bibinfo {volume} {79}},\ \bibinfo
  {pages} {073201} (\bibinfo {year} {2010})}\BibitemShut {NoStop}%
\bibitem [{\citenamefont {Suzuki}\ \emph {et~al.}(2013)\citenamefont {Suzuki}
  \emph {et~al.}}]{Suzuki13}%
  \BibitemOpen
  \bibfield  {author} {\bibinfo {author} {\bibfnamefont {H.}~\bibnamefont
  {Suzuki}} \emph {et~al.},\ }\href@noop {} {\bibfield  {journal} {\bibinfo
  {journal} {Nucl. Instr. and Meth. in Phys. Res. B}\ }\textbf {\bibinfo
  {volume} {317}},\ \bibinfo {pages} {756} (\bibinfo {year}
  {2013})}\BibitemShut {NoStop}%
\bibitem [{\citenamefont {Fukuda}\ \emph {et~al.}(2018)\citenamefont {Fukuda}
  \emph {et~al.}}]{Fukuda18}%
  \BibitemOpen
  \bibfield  {author} {\bibinfo {author} {\bibfnamefont {N.}~\bibnamefont
  {Fukuda}} \emph {et~al.},\ }\href@noop {} {\bibfield  {journal} {\bibinfo
  {journal} {J. Phys. Soc. Jpn.}\ }\textbf {\bibinfo {volume} {87}},\ \bibinfo
  {pages} {014202} (\bibinfo {year} {2018})}\BibitemShut {NoStop}%
\bibitem [{\citenamefont {Caama{\~n}o}\ \emph {et~al.}(2013)\citenamefont
  {Caama{\~n}o} \emph {et~al.}}]{Caamano13}%
  \BibitemOpen
  \bibfield  {author} {\bibinfo {author} {\bibfnamefont {M.}~\bibnamefont
  {Caama{\~n}o}} \emph {et~al.},\ }\href@noop {} {\bibfield  {journal}
  {\bibinfo  {journal} {Phys. Rev. C}\ }\textbf {\bibinfo {volume} {88}},\
  \bibinfo {pages} {024605} (\bibinfo {year} {2013})}\BibitemShut {NoStop}%
\bibitem [{\citenamefont {Tarasov}\ \emph {et~al.}(2018)\citenamefont {Tarasov}
  \emph {et~al.}}]{OT-EPJA18}%
  \BibitemOpen
  \bibfield  {author} {\bibinfo {author} {\bibfnamefont {O.~B.}\ \bibnamefont
  {Tarasov}} \emph {et~al.},\ }\href@noop {} {\bibfield  {journal} {\bibinfo
  {journal} {Eur. Phys. J. A}\ }\textbf {\bibinfo {volume} {54}},\ \bibinfo
  {pages} {66} (\bibinfo {year} {2018})}\BibitemShut {NoStop}%
\bibitem [{\citenamefont {McGaughey}\ \emph {et~al.}(1985)\citenamefont
  {McGaughey} \emph {et~al.}}]{McGaughey85}%
  \BibitemOpen
  \bibfield  {author} {\bibinfo {author} {\bibfnamefont {P.~L.}\ \bibnamefont
  {McGaughey}} \emph {et~al.},\ }\href@noop {} {\bibfield  {journal} {\bibinfo
  {journal} {Phys. Rev. C}\ }\textbf {\bibinfo {volume} {31}},\ \bibinfo
  {pages} {896} (\bibinfo {year} {1985})}\BibitemShut {NoStop}%
\bibitem [{\citenamefont {III}\ \emph {et~al.}(2009)\citenamefont {III} \emph
  {et~al.}}]{Folden09}%
  \BibitemOpen
  \bibfield  {author} {\bibinfo {author} {\bibfnamefont {C.~M.~F.}\
  \bibnamefont {III}} \emph {et~al.},\ }\href@noop {} {\bibfield  {journal}
  {\bibinfo  {journal} {Phys. Rev. C}\ }\textbf {\bibinfo {volume} {79}},\
  \bibinfo {pages} {064318} (\bibinfo {year} {2009})}\BibitemShut {NoStop}%
\bibitem [{\citenamefont {Neufcourt}\ \emph {et~al.}(2020)\citenamefont
  {Neufcourt}, \citenamefont {Cao}, \citenamefont {Giuliani}, \citenamefont
  {Nazarewicz}, \citenamefont {Olsen},\ and\ \citenamefont
  {Tarasov}}]{Neufcourt20}%
  \BibitemOpen
  \bibfield  {author} {\bibinfo {author} {\bibfnamefont {L.}~\bibnamefont
  {Neufcourt}}, \bibinfo {author} {\bibfnamefont {Y.}~\bibnamefont {Cao}},
  \bibinfo {author} {\bibfnamefont {S.}~\bibnamefont {Giuliani}}, \bibinfo
  {author} {\bibfnamefont {W.}~\bibnamefont {Nazarewicz}}, \bibinfo {author}
  {\bibfnamefont {E.}~\bibnamefont {Olsen}},\ and\ \bibinfo {author}
  {\bibfnamefont {O.~B.}\ \bibnamefont {Tarasov}},\ }\href@noop {} {\bibfield
  {journal} {\bibinfo  {journal} {Phys. Rev. C}\ }\textbf {\bibinfo {volume}
  {101}},\ \bibinfo {pages} {044307} (\bibinfo {year} {2020})}\BibitemShut
  {NoStop}%
\bibitem [{\citenamefont {Glasmacher}\ \emph {et~al.}(2017)\citenamefont
  {Glasmacher} \emph {et~al.}}]{Glasmacher17}%
  \BibitemOpen
  \bibfield  {author} {\bibinfo {author} {\bibfnamefont {T.}~\bibnamefont
  {Glasmacher}} \emph {et~al.},\ }\href@noop {} {\bibfield  {journal} {\bibinfo
   {journal} {Nucl. Phys. News}\ }\textbf {\bibinfo {volume} {27}},\ \bibinfo
  {pages} {28} (\bibinfo {year} {2017})}\BibitemShut {NoStop}%
\bibitem [{\citenamefont {Sherrill}(2018)}]{Sherrill18}%
  \BibitemOpen
  \bibfield  {author} {\bibinfo {author} {\bibfnamefont {B.~M.}\ \bibnamefont
  {Sherrill}},\ }\href@noop {} {\bibfield  {journal} {\bibinfo  {journal} {EPJ
  Web of Conferences}\ }\textbf {\bibinfo {volume} {178}},\ \bibinfo {pages}
  {01001} (\bibinfo {year} {2018})}\BibitemShut {NoStop}%
\bibitem [{\citenamefont {Kurcewicz}\ \emph {et~al.}(2012)\citenamefont
  {Kurcewicz} \emph {et~al.}}]{Kurcewicz12}%
  \BibitemOpen
  \bibfield  {author} {\bibinfo {author} {\bibfnamefont {J.}~\bibnamefont
  {Kurcewicz}} \emph {et~al.},\ }\href@noop {} {\bibfield  {journal} {\bibinfo
  {journal} {Physics Letter B}\ }\textbf {\bibinfo {volume} {717}},\ \bibinfo
  {pages} {371} (\bibinfo {year} {2012})}\BibitemShut {NoStop}%
\bibitem [{\citenamefont {Gaimard}\ and\ \citenamefont
  {Schmidt}(1991)}]{Gaimard91}%
  \BibitemOpen
  \bibfield  {author} {\bibinfo {author} {\bibfnamefont {J.-J.}\ \bibnamefont
  {Gaimard}}\ and\ \bibinfo {author} {\bibfnamefont {K.-H.}\ \bibnamefont
  {Schmidt}},\ }\href@noop {} {\bibfield  {journal} {\bibinfo  {journal} {Nucl.
  Phys. A}\ }\textbf {\bibinfo {volume} {531}},\ \bibinfo {pages} {709}
  (\bibinfo {year} {1991})}\BibitemShut {NoStop}%
\bibitem [{\citenamefont {Keli{\'c}}\ \emph {et~al.}(2008)\citenamefont
  {Keli{\'c}}, \citenamefont {Ricciardi},\ and\ \citenamefont
  {Schmidt}}]{ABLA07}%
  \BibitemOpen
  \bibfield  {author} {\bibinfo {author} {\bibfnamefont {A.}~\bibnamefont
  {Keli{\'c}}}, \bibinfo {author} {\bibfnamefont {M.~V.}\ \bibnamefont
  {Ricciardi}},\ and\ \bibinfo {author} {\bibfnamefont {K.-H.}\ \bibnamefont
  {Schmidt}},\ }in\ \href@noop {} {\emph {\bibinfo {booktitle} {Proceedings of
  the Joint ICTP-IAEA Advanced Workshop on Model Codes for Spallation
  Reactions, ICTP}}}\ (\bibinfo  {publisher} {IAEA INDC},\ \bibinfo {year}
  {2008})\ pp.\ \bibinfo {pages} {181--221},\ \bibinfo {note} {arXiv:0906.4193
  [nucl-th]}\BibitemShut {NoStop}%
\bibitem [{\citenamefont {Tarasov}(2005{\natexlab{a}})}]{LISE_AF_preprint}%
  \BibitemOpen
  \bibfield  {author} {\bibinfo {author} {\bibfnamefont {O.~B.}\ \bibnamefont
  {Tarasov}},\ }\href@noop {} {\emph {\bibinfo {title} {MSUCL1300}}},\ \bibinfo
  {type} {Tech. Rep.}\ (\bibinfo  {institution} {NSCL, Michigan State
  University},\ \bibinfo {year} {2005})\ \bibinfo {note}
  {\url{https://lise.nscl.msu.edu/7_5/lise++_7_5.pdf}}\BibitemShut {NoStop}%
\bibitem [{\citenamefont {Tarasov}(2005{\natexlab{b}})}]{LISE_AF}%
  \BibitemOpen
  \bibfield  {author} {\bibinfo {author} {\bibfnamefont {O.~B.}\ \bibnamefont
  {Tarasov}},\ }\href@noop {} {\bibfield  {journal} {\bibinfo  {journal} {Eur.
  Phys. J. A}\ }\textbf {\bibinfo {volume} {25}},\ \bibinfo {pages} {751}
  (\bibinfo {year} {2005}{\natexlab{b}})}\BibitemShut {NoStop}%
\bibitem [{\citenamefont {Tarasov}\ and\ \citenamefont
  {Bazin}(2008)}]{Tarasov08}%
  \BibitemOpen
  \bibfield  {author} {\bibinfo {author} {\bibfnamefont {O.~B.}\ \bibnamefont
  {Tarasov}}\ and\ \bibinfo {author} {\bibfnamefont {D.}~\bibnamefont
  {Bazin}},\ }\href@noop {} {\bibfield  {journal} {\bibinfo  {journal} {Nucl.
  Instr. and Meth. B}\ }\textbf {\bibinfo {volume} {266}},\ \bibinfo {pages}
  {4657} (\bibinfo {year} {2008})},\ \bibinfo {note}
  {\url{https://lise.nscl.msu.edu}}\BibitemShut {NoStop}%
\bibitem [{\citenamefont {Bazin}\ and\ \citenamefont
  {Sherrill}(1994)}]{Bazin94}%
  \BibitemOpen
  \bibfield  {author} {\bibinfo {author} {\bibfnamefont {D.}~\bibnamefont
  {Bazin}}\ and\ \bibinfo {author} {\bibfnamefont {B.~M.}\ \bibnamefont
  {Sherrill}},\ }\href@noop {} {\bibfield  {journal} {\bibinfo  {journal}
  {Phys. Rev. E}\ }\textbf {\bibinfo {volume} {50}},\ \bibinfo {pages} {4017}
  (\bibinfo {year} {1994})}\BibitemShut {NoStop}%
\bibitem [{\citenamefont {de~Jong}\ \emph {et~al.}(1997)\citenamefont
  {de~Jong}, \citenamefont {Ignatyuk},\ and\ \citenamefont
  {Schmidt}}]{deJong97}%
  \BibitemOpen
  \bibfield  {author} {\bibinfo {author} {\bibfnamefont {M.}~\bibnamefont
  {de~Jong}}, \bibinfo {author} {\bibfnamefont {A.}~\bibnamefont {Ignatyuk}},\
  and\ \bibinfo {author} {\bibfnamefont {K.-H.}\ \bibnamefont {Schmidt}},\
  }\href@noop {} {\bibfield  {journal} {\bibinfo  {journal} {Nucl. Phys. A}\
  }\textbf {\bibinfo {volume} {613}},\ \bibinfo {pages} {435} (\bibinfo {year}
  {1997})}\BibitemShut {NoStop}%
\bibitem [{\citenamefont {Bazin}\ \emph {et~al.}(2003)\citenamefont {Bazin}
  \emph {et~al.}}]{Bazin03}%
  \BibitemOpen
  \bibfield  {author} {\bibinfo {author} {\bibfnamefont {D.}~\bibnamefont
  {Bazin}} \emph {et~al.},\ }\href@noop {} {\bibfield  {journal} {\bibinfo
  {journal} {Nucl. Instr. and Meth. in Phys. Res. B}\ }\textbf {\bibinfo
  {volume} {204}},\ \bibinfo {pages} {629} (\bibinfo {year}
  {2003})}\BibitemShut {NoStop}%
\bibitem [{\citenamefont {Paschalis}\ \emph {et~al.}(2013)\citenamefont
  {Paschalis} \emph {et~al.}}]{Paschalis13}%
  \BibitemOpen
  \bibfield  {author} {\bibinfo {author} {\bibfnamefont {S.}~\bibnamefont
  {Paschalis}} \emph {et~al.},\ }\href@noop {} {\bibfield  {journal} {\bibinfo
  {journal} {Nucl. Instr. and Meth. in Phys. Res. A}\ }\textbf {\bibinfo
  {volume} {709}},\ \bibinfo {pages} {44} (\bibinfo {year} {2013})}\BibitemShut
  {NoStop}%
\bibitem [{\citenamefont {Stolz}\ \emph {et~al.}(2006)\citenamefont {Stolz}
  \emph {et~al.}}]{Stolz06}%
  \BibitemOpen
  \bibfield  {author} {\bibinfo {author} {\bibfnamefont {A.}~\bibnamefont
  {Stolz}} \emph {et~al.},\ }\href@noop {} {\bibfield  {journal} {\bibinfo
  {journal} {Diamond and Related Materials}\ }\textbf {\bibinfo {volume}
  {15}},\ \bibinfo {pages} {807} (\bibinfo {year} {2006})}\BibitemShut
  {NoStop}%
\bibitem [{\citenamefont {Michimasa}\ \emph {et~al.}(2013)\citenamefont
  {Michimasa} \emph {et~al.}}]{Michimasa13}%
  \BibitemOpen
  \bibfield  {author} {\bibinfo {author} {\bibfnamefont {S.}~\bibnamefont
  {Michimasa}} \emph {et~al.},\ }\href@noop {} {\bibfield  {journal} {\bibinfo
  {journal} {Nucl. Instr. and Meth. in Phys. Res. B}\ }\textbf {\bibinfo
  {volume} {317}},\ \bibinfo {pages} {710} (\bibinfo {year}
  {2013})}\BibitemShut {NoStop}%
\bibitem [{\citenamefont {Tarasov}\ \emph {et~al.}(2010)\citenamefont {Tarasov}
  \emph {et~al.}}]{OT-NIMA09}%
  \BibitemOpen
  \bibfield  {author} {\bibinfo {author} {\bibfnamefont {O.~B.}\ \bibnamefont
  {Tarasov}} \emph {et~al.},\ }\href@noop {} {\bibfield  {journal} {\bibinfo
  {journal} {Nucl. Instr. and Meth. in Phys. Res. A}\ }\textbf {\bibinfo
  {volume} {620}},\ \bibinfo {pages} {578} (\bibinfo {year}
  {2010})}\BibitemShut {NoStop}%
\bibitem [{\citenamefont {Faust}(2002)}]{Faust02}%
  \BibitemOpen
  \bibfield  {author} {\bibinfo {author} {\bibfnamefont {H.~R.}\ \bibnamefont
  {Faust}},\ }\href@noop {} {\bibfield  {journal} {\bibinfo  {journal} {Eur.
  Phys. J. A}\ }\textbf {\bibinfo {volume} {14}},\ \bibinfo {pages} {459}
  (\bibinfo {year} {2002})}\BibitemShut {NoStop}%
\bibitem [{\citenamefont {Tarasov}(2016)}]{LISE_Reverse_doc}%
  \BibitemOpen
  \bibfield  {author} {\bibinfo {author} {\bibfnamefont {O.~B.}\ \bibnamefont
  {Tarasov}},\ }\href@noop {} {\emph {\bibinfo {title} {Reverse configurations
  in \lisepp}}},\ \bibinfo {type} {Tech. Rep.}\ (\bibinfo  {institution} {NSCL,
  Michigan State University},\ \bibinfo {year} {2016})\ \bibinfo {note}
  {\url{https://lise.nscl.msu.edu/9_10/ReverseConfiguration.pdf}}\BibitemShut
  {NoStop}%
\bibitem [{\citenamefont {Makino}\ and\ \citenamefont {Berz}(2006)}]{Makino06}%
  \BibitemOpen
  \bibfield  {author} {\bibinfo {author} {\bibfnamefont {K.}~\bibnamefont
  {Makino}}\ and\ \bibinfo {author} {\bibfnamefont {M.}~\bibnamefont {Berz}},\
  }\href@noop {} {\bibfield  {journal} {\bibinfo  {journal} {Nucl. Instr. and
  Meth. in Phys. Res. A}\ }\textbf {\bibinfo {volume} {558}},\ \bibinfo {pages}
  {346} (\bibinfo {year} {2006})}\BibitemShut {NoStop}%
\bibitem [{\citenamefont {Scheidenberger}\ \emph {et~al.}(1998)\citenamefont
  {Scheidenberger} \emph {et~al.}}]{GLOBAL}%
  \BibitemOpen
  \bibfield  {author} {\bibinfo {author} {\bibfnamefont {C.}~\bibnamefont
  {Scheidenberger}} \emph {et~al.},\ }\href@noop {} {\bibfield  {journal}
  {\bibinfo  {journal} {Nucl. Instr. and Meth. in Phys. Res. B}\ }\textbf
  {\bibinfo {volume} {142}},\ \bibinfo {pages} {441} (\bibinfo {year}
  {1998})}\BibitemShut {NoStop}%
\bibitem [{\citenamefont {Lamour}\ \emph {et~al.}(2015)\citenamefont {Lamour}
  \emph {et~al.}}]{ETACHA4}%
  \BibitemOpen
  \bibfield  {author} {\bibinfo {author} {\bibfnamefont {E.}~\bibnamefont
  {Lamour}} \emph {et~al.},\ }\href@noop {} {\bibfield  {journal} {\bibinfo
  {journal} {Phys. Rev. A}\ }\textbf {\bibinfo {volume} {92}},\ \bibinfo
  {pages} {042703} (\bibinfo {year} {2015})}\BibitemShut {NoStop}%
\bibitem [{\citenamefont {Tarasov}\ \emph {et~al.}(2023)\citenamefont {Tarasov}
  \emph {et~al.}}]{LISE2023}%
  \BibitemOpen
  \bibfield  {author} {\bibinfo {author} {\bibfnamefont {O.~B.}\ \bibnamefont
  {Tarasov}} \emph {et~al.},\ }\href
  {https://doi.org/https://doi.org/10.1016/j.nimb.2023.04.039} {\bibfield
  {journal} {\bibinfo  {journal} {Nucl. Instr. and Meth. in Phys. Res. B}\
  }\textbf {\bibinfo {volume} {541}},\ \bibinfo {pages} {4} (\bibinfo {year}
  {2023})}\BibitemShut {NoStop}%
\bibitem [{\citenamefont {Winger}\ \emph {et~al.}(1992)\citenamefont {Winger}
  \emph {et~al.}}]{Intensity}%
  \BibitemOpen
  \bibfield  {author} {\bibinfo {author} {\bibfnamefont {J.~A.}\ \bibnamefont
  {Winger}} \emph {et~al.},\ }\href@noop {} {\bibfield  {journal} {\bibinfo
  {journal} {Nucl. Instr. and Meth. in Phys. Res. B}\ }\textbf {\bibinfo
  {volume} {70}},\ \bibinfo {pages} {380} (\bibinfo {year} {1992})}\BibitemShut
  {NoStop}%
\bibitem [{\citenamefont {Noji}\ \emph {et~al.}(2023)\citenamefont {Noji} \emph
  {et~al.}}]{HRS}%
  \BibitemOpen
  \bibfield  {author} {\bibinfo {author} {\bibfnamefont {S.}~\bibnamefont
  {Noji}} \emph {et~al.},\ }\href
  {https://doi.org/https://doi.org/10.1016/j.nima.2022.167548} {\bibfield
  {journal} {\bibinfo  {journal} {Nucl. Instr. and Meth. in Phys. Res. A}\
  }\textbf {\bibinfo {volume} {1045}},\ \bibinfo {pages} {167548} (\bibinfo
  {year} {2023})}\BibitemShut {NoStop}%
\end{thebibliography}%
\end{document}